\documentclass[12pt]{article}
\usepackage{epsfig}
\usepackage{amssymb}
\usepackage{amsmath}
\usepackage{amsfonts}
\usepackage{graphicx}
\usepackage{mathrsfs}
\usepackage{mathabx}
\DeclareMathAlphabet{\mathscrbf}{OMS}{mdugm}{b}{n}
\usepackage[dvips]{color}
\usepackage{multirow}
\usepackage{calc}
\usepackage{accents}

\makeatletter
\newcommand{\shorteq}{\mathrel{\mkern0.2mu\mathpalette\shorteq@\relax\mkern0.2mu}}
\newcommand{\shorteq@}[2]{\scalebox{0.5}[1]{$\m@th#1=$}}

\newcommand{\longeq}[1]{\mathrel{\mathpalette\longeq@{#1}}}
\newcommand{\longeq@}[2]{%
  \begingroup
  \sbox\z@{$\m@th#1=$}%
  \ifdim#2<\wd\z@
    \resizebox{#2}{\height}{\box\z@}%
  \else
    \ifdim#2<3\wd\z@
      \hbox to #2{$\m@th#1=\hss=\hss=\hss=$}%
    \else
      \hbox to #2{$\m@th#1=\cleaders\hbox to 0.2\wd\z@{\hss$#1=$\hss}\hfil=$}%
    \fi
  \fi
  \endgroup
}
\makeatother

\newcommand{\bsigma}{\boldsymbol{\sigma}}

\newcommand{\bnabla}{\boldsymbol{\nabla}}

\newcommand{\R}{\mathbb{R}}
\newcommand{\C}{\mathbb{C}}
\newcommand{\Z}{\mathbb{Z}}
\newcommand{\N}{\mathbb{N}}

\newcommand{\fa}{\mathfrak{a}}
\newcommand{\fb}{\mathfrak{b}}
\newcommand{\fc}{\mathfrak{c}}

\newcommand{\ff}{\mathfrak{f}}

\newcommand{\fm}{\mathfrak{m}}
\newcommand{\fn}{{\mathfrak{n}}}

\newcommand{\fs}{\mathfrak{s}}

\newcommand{\fK}{\mathfrak{K}}

\newcommand{\fM}{\mathfrak{M}}

\newcommand{\bfe}{\mathbf{e}}

\newcommand{\bm}{\mathbf{m}}

\newcommand{\bfr}{\mathbf{r}}

\newcommand{\bC}{\mathbf{C}}

\newcommand{\bcE}{{\boldsymbol{\cE}}}

\newcommand{\bG}{\mathbf{G}}
\newcommand{\bH}{\mathbf{H}}
\newcommand{\bI}{\mathbf{I}}

\newcommand{\bL}{\mathbf{L}}

\newcommand{\bM}{\mathbf{M}}

\newcommand{\bS}{\mathbf{S}}

\newcommand{\cA}{{\mathcal{A}}}
\newcommand{\cB}{\mathcal{B}}

\newcommand{\cH}{\mathcal{H}}
\newcommand{\cE}{\mathcal{E}}

\newcommand{\cK}{\mathcal{K}}

\newcommand{\cN}{\mathcal{N}}

\newcommand{\cP}{\mathcal{P}}
\newcommand{\cQ}{\mathcal{Q}}
\newcommand{\cR}{\mathcal{R}}

\newcommand{\cT}{\mathcal{T}}
\newcommand{\cU}{\mathcal{U}}
\newcommand{\cV}{\mathcal{V}}
\newcommand{\cW}{\mathcal{W}}
\newcommand{\cX}{\mathcal{X}}
\newcommand{\cY}{\mathcal{Y}}

\newcommand{\be}{\begin{equation}}
\newcommand{\ee}{\end{equation}}
\newcommand{\bea}{\begin{eqnarray}}
\newcommand{\eea}{\end{eqnarray}}
\newcommand{\nn}{\nonumber}
\newcommand{\kt}{\rangle}
\newcommand{\br}{\langle}

\newcommand{\ed}{\end{document}}

\newcommand{\bi}{\begin{itemize}}
\newcommand{\ei}{\end{itemize}}

\newcommand{\bce}{\begin{center}}
\newcommand{\ece}{\end{center}}
\newcommand{\sA}{\mathscr{A}}

\newcommand{\sF}{\mathscr{F}}

\newcommand{\sH}{\mathscr{H}}
\newcommand{\bsH}{\mathscrbf{H}}

\newcommand{\sM}{\mathscr{M}}
\newcommand{\bsM}{\mathscrbf{M}}

\newcommand{\sN}{\mathscr{N}}

\newcommand{\sR}{\mathscr{R}}

\newcommand{\sT}{\mathscr{T}}
\newcommand{\sV}{\mathscr{V}}

\newcommand{\bsU}{\mathscrbf{U}}
\newcommand{\RE}{{\rm Re}}
\newcommand{\IM}{{\rm Im}}

\newcommand{\bcK}{{\boldsymbol{\cK}}}

\newcommand{\bcH}{{\boldsymbol{\cH}}}
\newcommand{\bcU}{{\boldsymbol{\cU}}}

\newcommand{\bzero}{{\boldsymbol{0}}}

\newcommand{\for}{{\mbox{\rm for}}}

\begin{document}

\title{Exactly Solvable Diffraction-Grating Scattering\\ Problems for 
%Transverse Electric and Transverse Magnetic 
TE and TM waves}

\author{Farhang Loran\thanks{E-mail address: loran@iut.ac.ir}~ and 
Ali~Mostafazadeh\thanks{Corresponding author, 
E-mail address: amostafazadeh@ku.edu.tr}\\[6pt]
$^*$Department of Physics, Isfahan University of Technology, \\ Isfahan 84156-83111, Iran\\[6pt]$^\dagger$
Departments of Mathematics and Physics, Ko\c{c} University,\\  34450 Sar{\i}yer,
Istanbul, T\"urkiye}

\date{ }
\maketitle

\begin{abstract}

In J.~Phys.~A {\bf 31}, 3493 (1998), Berry provides an analytic solution of %establishes the exact solvability of 
the problem of determining the diffracted beam intensities for atoms incident upon a grating given by the potential, \vspace{-6pt}
	\[v(x,y):=\left\{\begin{array}{cc}
	iV_0(e^{i\fK\,y}-1)&\for~~0\leq x\leq\ell,\\
	0 &{\rm otherwise},\end{array}\right.\]
%$v(x,y):=iV_0(e^{i\fK\,y}-1)$ for $x\in[0,\ell]$ and $v(x,y):=0$ for $x\notin[0,\ell]$, 
where $V_0, \fK$, and $\ell$ are positive real parameters. This result, which relies on the paraxial approximation, readily applies to the diffraction of transverse electric (TE) waves by the nonmagnetic optical grating given by the relative permittivity, $\hat\varepsilon(x,y):=1-v(x,y)/k^2$, where $k$ is the incident wavenumber. We show that Berry's grating belongs to a larger class of possibly magnetic diffraction gratings whose scattering problem for both TE and transverse magnetic (TM) waves is exactly solvable. Our analysis makes use of a recently developed dynamical formulation of stationary scattering that is based on the idea of mapping the scattering problem to the quantum dynamics generated by an effective non-Hermitian Hamiltonian operator. {We use this formulation to derive explicit analytic expressions for the diffracted beam amplitudes that are valid beyond paraxial approximation.} As a concrete example, we offer the exact solution of the scattering problem for TE and TM waves incident upon a generalization of Berry's grating. 

\end{abstract}

\section{Introduction} 
Exactly solvable models are indispensable tools for understanding the basic features and further development of physical theories. In scattering theory, they are as rare as in any other area of physics research. This makes the discovery of new exactly solvable scattering problems extremely valuable. The atomic diffraction grating examined by Berry in Ref.~\cite{berry-1998b} is a good example. It is equivalent to the problem of the transmission of a time-harmonic $z$-polarized transverse electric (TE) wave through a nonmagnetic optical grating with relative permittivity 
	\be
	\hat\varepsilon(x,y):=1-\frac{v(x,y)}{k^2}=
	1+\frac{iV_0}{k^2}
	\left\{\begin{array}{cc}
	1-e^{i\fK\,y}&\for~0\leq x\leq \ell,\\
	0&{\rm otherwise},\end{array}\right.
	\label{berry1}
	\ee
where $v$ is the potential given in the abstract, and $k$ is the incident wavenumber \cite{born-wolf}. 

In Refs.~\cite{pra-2017}, we establish the exact solvability of the scattering problem defined by a potential of the form
	\be
	v(x,y):=\left\{\begin{array}{cc}
	v_0+v_1(x) e^{i\fK\,y}&\for~~0\leq x\leq \ell,\\
	0&{\rm otherwise},\end{array}\right.
	\label{exact-v}
	\ee
where $v_0$ is a real or complex constant, and $v_1$ is an arbitrary (piecewise continuous) possibly complex-valued function of $x$.\footnote{This scattering problem is also exactly solvable if $v_0$ depends on $x$ in such a way that the one-dimensional scattering problem defined by $v_0(x)$ is exactly solvable \cite{pra-2017}.} This is identical to the scattering problem for normally-incident TE waves scattered by a nonmagnetic optical grating corresponding to the permittivity profile,
	\be
	\hat\varepsilon(x,y)=1+
	\left\{\begin{array}{cc}
	\fa_0+\fa_1(x) e^{i\fK\,y}&\for~~0\leq x\leq \ell,\\
	0&{\rm otherwise},\end{array}\right.
	\label{exact-1}
	\ee
where $\fa_0:=-v_0/k^2$ and $\fa_1(x)=-v_1(x)/k^2$.  

It is clear that Berry's grating  \eqref{berry1}  is a special case of the nonmagnetic diffraction gratings given by \eqref{exact-1}, but the approaches pursued in Refs.~\cite{berry-1998b} and \cite{pra-2017} to study them are completely different. The analysis of Ref.~\cite{berry-1998b} uses the Raman-Nath formalism for treating diffraction gratings \cite{raman-nath,berry-1966} and relies on the observation that for Berry's grating the analog of the Raman-Nath difference equation for diffracted beam intensities is exactly solvable. In contrast, Ref.~\cite{pra-2017} employs a recently proposed dynamical formulation of stationary scattering (DFSS) whose basic idea is to map the scattering problem to the determination of the time-evolution operator for a corresponding effective non-Hermitian Hamiltonian operator \cite{ap-2014,pra-2021,pra-2023}. Because this Hamiltonian is non-stationary, its time-evolution operator does not generally admit a closed-form expression; it can only be expressed as a formal Dyson series. As a result, for a generic scattering problem, DFSS yields a series expansion of the scattering amplitude. The exact solvability of the scattering problem for potentials of the form \eqref{exact-v} stems from the fact that the corresponding series expansion of the scattering amplitude truncates \cite{pra-2017,pra-2021}. 

It must also be noted that the Raman-Nath formalism and, in particular, the derivation of the analog of the Raman-Nath equation for Berry's grating given in \cite{berry-1998b} make use of paraxial approximation, which is applicable for small incidence angles. In contrast, the analysis of Ref.~\cite{pra-2017} applies to arbitrary incidence angles. It is valid beyond the Raman-Nath regime, i.e., it does not assume that $\fK/k$ or $k\ell$ is small \cite{Klein-1967,Moharam-1978}.  

The progress towards an extension of the results of Refs.~\cite{berry-1998b,pra-2017} to TM waves and magnetic diffraction gratings has been hindered by the fact that the corresponding scattering problems are defined not by the time-independent Schr\"odinger equation or equivalently Helmholtz equation, but by an equation of the form
	\be
	\alpha\,\vec\nabla\cdot(\alpha^{-1} \vec\nabla \psi)+k^2\fn^2\psi=0,
	\label{bergmann-eq-1}
	\ee
where $\alpha$ and $\fn$ are possibly complex-valued functions of the space coordinates \cite{Reutskiy}. Eq.~\eqref{bergmann-eq-1} is known as the Bergmann equation in acoustics; it describes the propagation of sound waves in a compressible fluid with mass density $\alpha$ and refractive index $\fn$, \cite{Bergmann,Martin}.\footnote{Refractive index of the fluid is, by definition, the ratio of the speed of sound at spatial infinity to that at a point inside the fluid \cite{Martin}.}
 
In Ref.~\cite{ptep-2024b}, we develop a DFSS for TE and TM waves scattered by the inhomogeneities of an effectively one-dimensional material, and in Ref~\cite{jpa-2025} we explore its application in constructing the low-frequency series expansions of the corresponding reflection and transmission amplitudes. Recently, we have extended the results of Refs.~\cite{ptep-2024b,jpa-2025} to the scattering problems defined by the Bergmann equation in two dimensions \cite{ptep-2026a}. {In particular, we outlined a DFSS for TE and TM waves scattered by an effectively two-dimensional material and explored its applications in the low-frequency scattering calculations and low-frequency invisibility. The purpose of the present article is to use the DFSS for TE and TM waves given in \cite{ptep-2026a}} to establish the exact solvability of the scattering problem for the TE and TM waves incident upon a diffraction grating whose relative permittivity and permeability are respectively given by
	\be
	\begin{aligned}
	&\hat\varepsilon(x,y)=1+\chi_\ell(x)\Big[\fa_0+\sum_{n=1}^N\fa_n(x)\, e^{in\fK\,y}\Big],\\
	&\hat\mu(x,y)=1+\chi_\ell(x)\Big[\fb_0+\sum_{n=1}^N\fb_n(x)\, e^{in\fK\,y}\Big],
	\end{aligned}
	\label{model-1}
	\ee
where
	\be
	\chi_\ell(x):=
	\left\{\begin{array}{cc}
	1&\for~0\leq x\leq \ell,\\
	0&{\rm otherwise},\end{array}\right.
	\label{chi}
	\ee
$\fa_0$ and $\fb_0$ are real or complex constants, $\ell$ and $\fK$ are positive real parameters, $N$ is a positive integer, and $\fa_n$ and $\fb_n$ with $n\geq1$ are real- or complex-valued functions of $x$. This diffraction grating is clearly a generalization of the nonmagnetic diffraction gratings given by \eqref{exact-1} and, in particular, Berry's grating \eqref{berry1}. {By its exact solvability, we mean that there is a procedure for analytic computation of its scattering amplitude (and diffracted beam intensities) that involves performing finitely many algebraic operations and definite integrals while avoiding any kind of approximation.}

The outline of the rest of this article is as follows. In Sec.~\ref{S2}, we present a basic  description of the scattering of TE and TM waves by a diffraction grating and offer a concise review of its dynamical formulation {as presented in Ref.~\cite{ptep-2026a}.} In Sec.~\ref{S3}, we examine the application of this formulation for the diffraction gratings given by \eqref{model-1} and establish the exact solvability of the corresponding scattering problem for both TE and TM waves. {Here we outline a method for analytic calculation of diffracted beam amplitudes, apply it to derive explicit analytic expressions for the amplitudes of the zeroth- and first-order diffracted beams, and provide an independent check on the validity of these expressions by showing that they fulfill the stringent requirements of the reciprocity principle.} In Sec.~\ref{S4}, we confine our attention to the scattering of TE and TM waves by a nonmagnetic grating with a relative permittivity of the form \eqref{exact-1}. In Sec.~\ref{S5}, we summarize our main findings and present our concluding remarks. {Accompanying appendices include the derivations of some of the results we employ in the main text.}
  
 \section{Scattering of TE and TM waves by a diffraction grating}
 \label{S2}
  
\subsection{TE and TM waves in an effectively 2D material}

Consider the propagation of time-harmonic electromagnetic waves in a linear stationary material with permittivity $\varepsilon$ and permeability $\mu$. Let $\bfr$ denote the position vector with Cartesian coordinates $(x,y,z)$, and suppose that as $r:=|\bfr|\to\infty$, $\varepsilon(\bfr)$ and $\mu(\bfr)$ tend to constant values $\varepsilon_{\rm b}$ and $\mu_{\rm b}$ at such a rate that the bounded solutions of the wave equation tend to superpositions of plane waves as $r\to\infty$. In the absence of free charges and currents,  Maxwell's equations for this system take the form
	\begin{align}
	&\bnabla\cdot(\hat\varepsilon\,\bcE)=0,
	&& \bnabla\cdot(\hat\mu\,\bcH)=0,
	\nn\\% \label{Max-1}\\
	&\bnabla\times\bcE=ik\hat\mu\,\bcH,
	&&\bnabla\times\bcH=-ik\hat\varepsilon\,\bcE,
    	\nn%\label{Max-2} 
	\end{align}
where $\bcE$ and $\bcH$ are vector-valued functions of $\bfr$ that respectively determine the electric and magnetic fields of the wave according to
$e^{-i\omega t}\bcE(\bfr)/\sqrt{\varepsilon_{\rm b}}$ and $e^{-i\omega t}\bcH(\bfr)/\sqrt{\mu_{\rm b}}$, $\omega$ is the angular frequency of the wave, $\hat\varepsilon:=\varepsilon/\varepsilon_{\rm b}$ and $\hat\mu:=\mu/\mu_{\rm b}$ are respectively the relative permittivity and permeability of the material, $k:=\omega/c_{{\rm b}}$,  and $c_{\rm b}:=1/\sqrt{\varepsilon_{\rm b}\mu_{\rm b}}$.~{\footnote{{We use the standard convention where $i:=\sqrt{-1}$, and the material has regions of loss (respectively gain) if the imaginary parts of both of $\hat\varepsilon$ and $\hat\mu$ can take positive (respectively negative) values or the imaginary part of one of them can take positive (respectively negative) values while the other takes real values.}}}

Suppose that $\varepsilon$ and $\mu$ do not depend on $z$. Then Maxwell's equations admit TE and TM wave solutions which respectively correspond to time-harmonic waves whose electric and magnetic fields point along the $z$ axis.\footnote{We identify TE and TM waves with those whose electric and magnetic fields are respectively orthogonal to the plane of incidence \cite{born-wolf}, which is the $x$-$y$ plane of our coordinate system. This definition differs from the one that takes TE and TM waves to be those whose electric and magnetic fields are respectively orthogonal to the symmetry axis of the medium \cite{taflov}.} More specifically,  
	\begin{align}
	&\mbox{for TE waves}: \left\{\begin{aligned}
	&\cE_x=\cE_y=0,~~\cE_z= \psi,\\
	&\cH_z=\partial_z\cH_x=\partial_z\cH_y=0,\\
	&\partial_x\cH_y-\partial_y\cH_x=-ik\,\hat\varepsilon\, \psi,
	\end{aligned}\right.
	%\nn\\[6pt]%\label{TE-def-2}
	&&\mbox{for TM waves}: \left\{\begin{aligned}
	&\cH_x=\cH_y=0,~~\cH_z= \psi,\\
	&\cE_z=\partial_z\cE_x=\partial_z\cE_y=0,\\
	&\partial_x\cE_y-\partial_y\cE_x=ik\,\hat\mu\, \psi,
	\end{aligned}~\right.
	\label{TE-TM-def}
	\end{align}
where $\cE_u$ and $\cH_u$ respectively stand for the $u$ components of $\bcE$ and $\bcH$, $u\in\{x,y,z\}$, $\psi$ is a possibly complex-valued function of $x$ and $y$ that satisfies 
	\begin{align}
	&\vec\nabla\cdot(\alpha^{-1} \vec\nabla \psi)+k^2\beta\,\psi=0,
	\label{Bergmann}
	\end{align}
$\vec\nabla:=\hat\bfe_x\partial_x+\hat\bfe_y\partial_y$, {$\hat\bfe_u$ stands for the unit vector pointing along the positive $u$ axis,} and $\alpha$ and $\beta$ are functions of $x$ and $y$ that are given by
	\begin{align}
	&\alpha:=\left\{\begin{array}{ccc}
	\hat\mu&\for&\mbox{TE waves},\\
	\hat\varepsilon&\for&\mbox{TM waves},
	\end{array}\right.
	&&\beta:=\left\{\begin{array}{ccc}
	\hat\varepsilon&\for&\mbox{TE waves},\\
	\hat\mu&\for&\mbox{TM waves}.
	\end{array}\right.
	\label{alpha-beta}
	\end{align}
	
Multiplying both sides of \eqref{Bergmann} by $\alpha$ and noting that the complex index of refraction $\fn$ of the medium satisfies 	
	\be
	\fn^2=\hat\varepsilon\hat\mu=\alpha\beta,
	\label{n2=}
	\ee
we can identify \eqref{Bergmann} with the Bergmann equation~\eqref{bergmann-eq-1}. The scattering of TE and TM waves due to the inhomogeneities of the medium corresponds to the solutions of this equation that fulfill 
	\be
	\psi(\vec r)\to
	\frac{{\sA_0}}{2\pi}\left[e^{i\vec k_0\cdot\vec r}+\sqrt{\frac{i}{ k r}}\,e^{ikr}\,\ff(\theta)\right]~~~\for~~~r\to\infty,
	\label{scattering}
	\ee
where {$\sA_0$ is a constant parameter representing the complex amplitude of the incident wave,} \linebreak $\vec r:=x\,\hat\bfe_x+y\,\hat\bfe_y$, $\vec k_0$ is the incident wave vector, $r$ and $\theta$ are polar coordinates of $\vec r$, and $\ff$ is the scattering amplitude \cite{adhikari-86}. The latter is a function of $\theta$ that depends on the incident wavenumber $k$ and the direction of $\vec k_0$ which we quantify using the angle $\theta_0$ between $\vec k_0$ and the positive $x$ axis.  $\theta_0$ is called the incidence angle. The $x$ component of $\vec k_0$ is given by $k_{0x}=k\cos\theta_0$.

\subsection{Dynamical formulation of the scattering of TE and TM waves by a diffraction grating}

Consider the relative permittivity and permeability profiles of the form  
	\begin{align}
	&\hat\varepsilon(x,y)=1+\chi_\ell(x)\fa(x,y),
	&&\hat\mu(x,y)=1+\chi_\ell(x)\fb(x,y),
	\label{slab}
	\end{align}
where $\chi_\ell$ is given by \eqref{chi}, and $\fa$ and $\fb$ are piecewise continuous functions of $x$ and $y$. Eq.~\eqref{slab} describes an infinite planar slab of thickness $\ell$ that is placed inside a homogeneous background with permittivity $\varepsilon_{\rm b}$ and permeability $\mu_{\rm b}$. It corresponds to a diffraction grating, if $\fa$ and $\fb$ are periodic functions of $y$. For example, for the gratings given by \eqref{model-1}, they have the form
	\begin{align}
	&\fa(x,y)=\fa_0+\sum_{n=1}^N\fa_n(x)e^{in\fK\,y},
	&&\fb(x,y)=\fb_0+\sum_{n=1}^N\fb_n(x)e^{in\fK\,y}.\nn
	\end{align}

\begin{figure}
        \begin{center}
        \includegraphics[scale=.25]{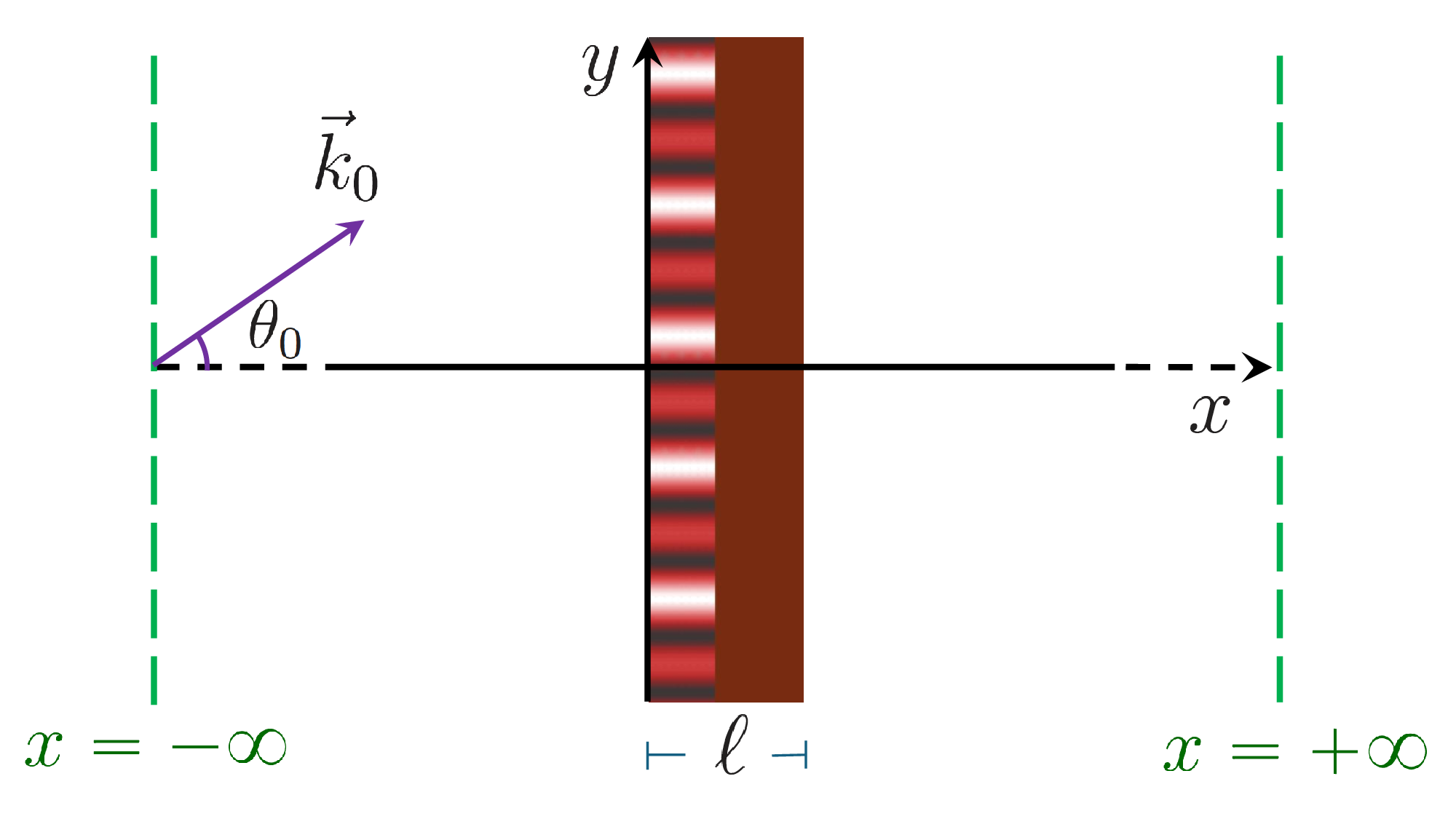}~~~~ 
        \includegraphics[scale=.25]{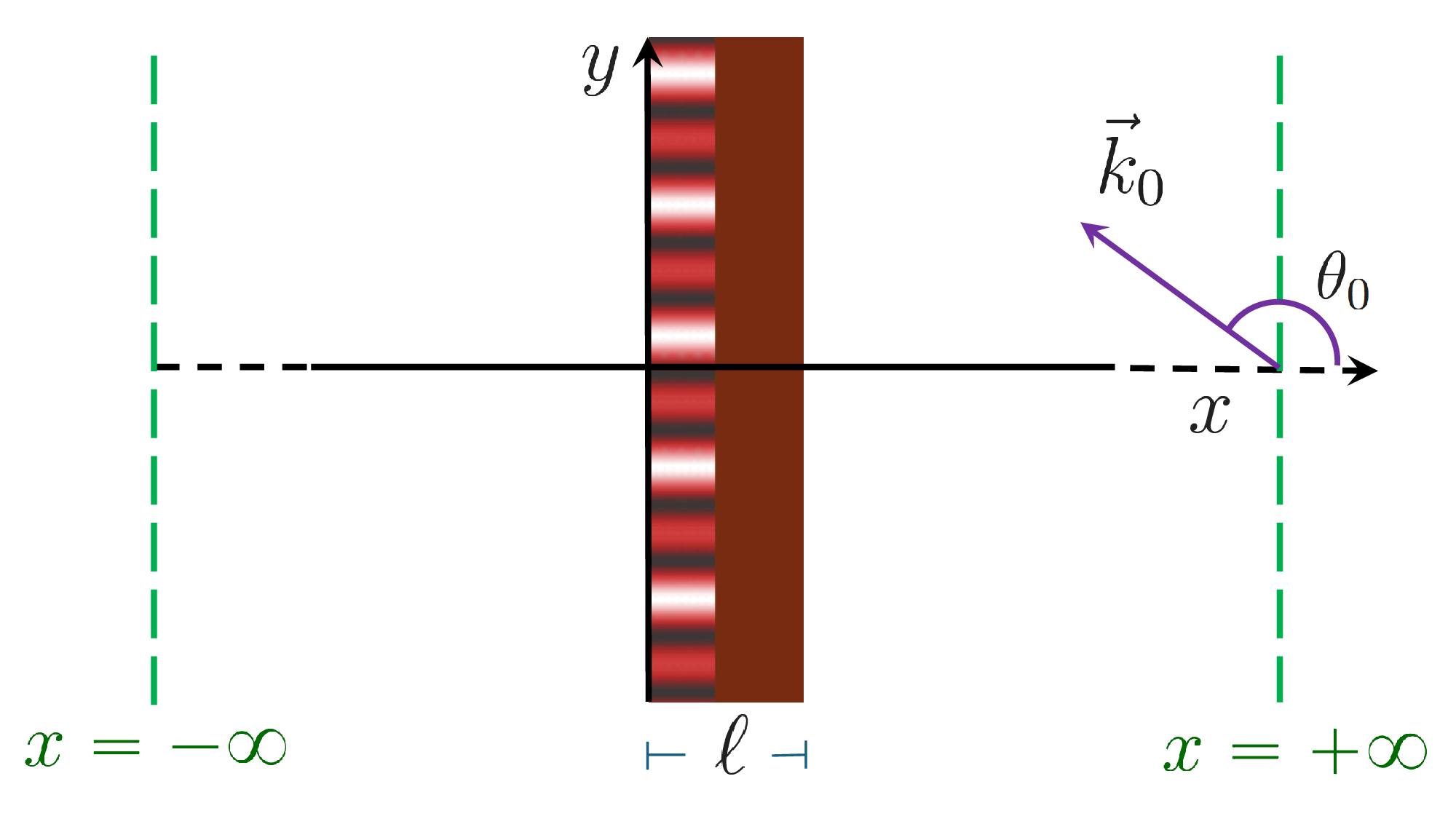} 
        \caption{Schematic views of the scattering setup for the scattering of left- and right-incident TE or TM waves (respectively on the left and right) by a diffraction grating of thickness $\ell$. The green dashed lines correspond to $x=\pm\infty$ where the detectors are located.}
        \label{fig1}
        \end{center}
        \end{figure}

The scattering of TE and TM waves by a grating given by \eqref{slab} is meaningful only if the incident wave vector $\vec k_0$ points towards the grating, i.e., $k_{0x}\neq 0$. If $k_{0x}>0$, $\cos\theta_0>0$, the incidence angle $\theta_0$ ranges over the interval $(-90^\circ,90^\circ)$, and we speak of the scattering of a left-incident wave. If $k_{0x}<0$, $\cos\theta_0<0$, $\theta_0$ ranges over $(90^\circ,270^\circ)$, and we speak of a right-incident wave. See Fig.~\ref{fig1}.

Next, we recall that the $\vec r$ appearing in \eqref{scattering} denotes the position of a generic detector that measures the scattered wave. Because $(r,\theta)$ are the polar coordinates of $\vec r$, the condition $r\to\infty$ in \eqref{scattering} reflects the fact that the detectors must be placed far away from the interaction region. For a diffraction grating given by \eqref{slab}, this means that they cannot be placed on the $y$ axis, i.e., $\cos\theta\neq 0$. Because $x=r\cos\theta$, this shows that the condition $r\to\infty$ in \eqref{scattering} is equivalent to $x\to\pm\infty$. Therefore, without loss of generality, we can assume that the detectors are mounted on the lines $x=\pm\infty$. Figure~\ref{fig1} provides a schematic view of the scattering setup for the left- and right-incident waves scattered by a diffraction grating.
 
According to \eqref{alpha-beta} and \eqref{slab}, for $x<0$ and $x>\ell$, $\hat\varepsilon=\hat\mu=1$, $\hat\alpha=\hat\beta=1$, and Bergmann's equation \eqref{Bergmann} reduces to the Helmholtz equation $\vec\nabla^2\psi+k^2\psi=0$. This implies that the bounded solutions of \eqref{Bergmann} tend to superpositions of plane waves as $x\to\pm\infty$. In other words, there are functions\footnote{We use the terms ``function'' and ``tempered distribution'' synonymously.} $A_\pm$ and $B_\pm$ such that 
	\be
	\psi(x,y)\to{\sA_0}\int_{-k}^k \frac{dp}{4\pi^2\varpi(p)}
	\Big[A_\pm(p) e^{i\varpi(p)x}+B_\pm(p) e^{-i\varpi(p)x}\Big]e^{ipy}~~~~\for~~~~x\to\pm\infty,
	\label{asym}
	\ee
where
	\be	
	\varpi(p):=\left\{\begin{array}{ccc}
	\sqrt{k^2-p^2}&\for&|p|<k,\\
	i\sqrt{p^2-k^2}&\for&|p|\geq k.\end{array}\right.
	\label{varpi-def}
	\ee
To be able to view the integral in \eqref{asym} as a Fourier integral, we identify $A_\pm$ and $B_\pm$ with functions defined on $\R$ by demanding that $A(p)=B(p)=0$ for $|p|\geq k$. This turns $A_\pm$ and $B_\pm$ into elements of 
	\[\sF_k:=\{\,f\in\sF\,|\,f(p)=0~\for~|p|\geq k\,\},\]
where $\sF$ stands for the vector space of (complex-valued) functions  defined on $\R$ that possess Fourier transform (more generally tempered distributions).

According to \eqref{asym}, $A_\pm$ and $B_\pm$ are the Fourier modes of the right-going and left-going waves with respect to the standard orientation of the $x$ axis. For a left-incident wave, $A_-$ must produce the wave function for the incident wave, i.e., the first term on the right-hand side of \eqref{scattering}, and $B_+$ must vanish, because there is no left-going wave at $x=+\infty$. Similarly, for a right-incident wave $B_+$ must produce the (left-going) incident wave, and $A_-=0$, because there is no right-going wave at $x=-\infty$. Using the superscripts $l$ and $r$ to identify the $A_\pm$ and $B_\pm$ corresponding to the left- and right-incident waves, we can infer from these observations and \eqref{asym} that
	\begin{align}
	&A_-^l(p)=B_+^r(p)=\check\delta_{p_0}(p),
	&&B_+^l(p)=A_-^r(p)=0,
	\label{LR-waves}
	\end{align}
where 
	\begin{align}
	&\check\delta_{p_0}(p):=2\pi\varpi_0\:\delta(p-p_0),
	&&\varpi_0:=\varpi(p_0)=k|\cos\theta_0|,
	&&p_0:=k\sin\theta_0,
	\label{tdelta}
	\end{align}
and $\delta(\cdot)$ stands for the Dirac delta function. In view of \eqref{scattering}, \eqref{asym}, and \eqref{LR-waves}, the scattering amplitude $\ff$ must be uniquely determined by $A^{l/r}_+$ and $B^{l/r}_-$. As we show in Ref.~\cite{pra-2021}, these equations imply
	\begin{align}
	\ff(\theta)=\frac{-i}{\sqrt{2\pi}}\times\left\{
	\begin{array}{ccc}
	A^l_+(k\sin\theta)-2\pi\delta(\theta-\theta_0)&\for&
	\cos\theta_0>0~\&~\cos\theta>0,\\
	B^l_-(k\sin\theta)&\for&
	\cos\theta_0>0~\&~\cos\theta<0,\\
	A^r_+(k\sin\theta)&\for&
	\cos\theta_0<0~\&~\cos\theta>0,\\
	B^r_-(k\sin\theta)-2\pi\delta(\theta-\theta_0)&\for&
	\cos\theta_0<0~\&~\cos\theta<0.
	\end{array}\right.
	\label{f=}
	\end{align}

Because \eqref{scattering}, \eqref{asym}, and \eqref{f=} are identical to those appearing in the discussion of the potential scattering in two dimensions \cite{pra-2021}, we can use the same analysis to define a fundamental transfer matrix $\widehat\bM$ for TE and TM waves scattered by a diffraction grating and express the corresponding scattering amplitude $\ff$ in terms of $\widehat\bM$. This is done in Ref.~\cite{ptep-2026a} for the more general situations where the scattering of TE and TM waves is due to the inhomogeneities of a two-dimensional medium which need not be confined to an infinite  planar slab. In the following, we provide a brief summary of the results of \cite{ptep-2026a} that we employ in this article.

The fundamental transfer matrix is the $2\times 2$ matrix $\widehat\bM$ that satisfies	\be
	\widehat\bM \left[\begin{array}{c}
	A_-\\
	B_-\end{array}\right]=\left[\begin{array}{c}
	A_+\\
	B_+\end{array}\right],
	\label{M-def}
	\ee
where $A_\pm$ and $B_\pm$ are the coefficient functions determining the asymptotic form of the solutions of Bergmann's equation \eqref{Bergmann},  {\cite{ptep-2026a}}. This is identical in form to the transfer matrix for TE and TM waves scattered by an effectively one-dimensional medium \cite{ptep-2024b} with the important difference that its entries, $\widehat M_{ab}$, are not numbers but linear operators acting in the function space $\sF_k$. 

Because $A_\pm$ and $B_\pm$ belong to $\sF_k$, the two-component functions appearing in \eqref{M-def} belong to the space $\sF_k^2$ of two-component functions of $p$ that evaluate to $0$ for $|p|\geq k$, i.e.,
	\be
	A_\pm,B_\pm\in\sF_k^2:=\{\,\phi\in\sF^2\,|\,\phi(p)=0~\for~|p|\geq k\,\},
	\label{AB-in}
	\ee
where $\sF^2$ stands for the space of two-component functions of $p$. According to \eqref{M-def} and \eqref{AB-in}, $\widehat\bM$ is a linear operator that maps $\sF_k^2$ to $\sF_k^2$. Therefore, it defines a linear operator acting $\sF^2_k$. In what follows, we use the symbol $\widehat\bM$ also for its restriction to $\sF_k^2$. The same applies for the entries of $\widehat\bM$, i.e., we view $\widehat M_{ab}$ as operators acting in $\sF_k$.
 
Substituting \eqref{LR-waves} in \eqref{M-def}, we find
	\begin{align}
	&A^l_+=\widehat M_{12}B^l_-+\widehat M_{11}\check\delta_{p_0},
	&&A^r_+=\widehat M_{12}B^r_-,
	\label{As=}\\
	&\widehat M_{22} B^l_-=-\widehat M_{21}\check\delta_{p_0},
	&&\widehat M_{22} B^r_-=\check\delta_{p_0}.
	\label{Bs=}
	\end{align}
In Ref.~\cite{pra-2025}, we use these equations to obtain series expansions for
$A^{l/r}_+$ and $B^{l/r}_-$. These involve positive integer powers of the operators $N_{ab}:\sF_k\to\sF_k$ given by
	\be
	\widehat N_{ab}:=\delta_{ab}\widehat I-\widehat M_{ab},
	\label{N-ab=}
	\ee
where $\delta_{ab}$ denotes the Kronecker delta symbol, and 
$\widehat I$ is the identity operator.\footnote{Throughout this article, we use $\widehat I$ and $\widehat 0$ to respectively denote the identity and zero operators acting in the relevant function spaces.} In particular, we have {\cite{ptep-2026a}},
	\begin{align}
	&B^l_-(k\sin\theta)=2\pi k|\cos\theta_0|
	\sum_{j=0}^\infty\br p_1|\widehat N_{22}^j \widehat N_{21}|p_0\kt,
	\label{B-L-series} \\
	&B^r_-(k\sin\theta)-2\pi\delta(\theta-\theta_0)=2\pi k|\cos\theta_0|
	\sum_{j=1}^\infty\br p_1|\widehat N_{22}^j |p_0\kt,	
	%=2\pi \varpi(p_0)\sum_{j=0}^\infty\br p|\widehat N_{22}^j |p_0\kt ,
	\label{B-R-series}\\
	&A^l_+(k\sin\theta)-2\pi\delta(\theta-\theta_0)=
	-2\pi k|\cos\theta_0|\Big[
	\br p_1|\widehat N_{11}|p_0\kt+
	\sum_{j=0}^\infty
	\br p_1|\widehat N_{12} \widehat N_{22}^j\widehat N_{21}|p_0\kt\Big],
	\label{A-L-series} \\
	&A^r_+(k\sin\theta)=-2\pi k|\cos\theta_0|\sum_{j=0}^\infty\br p_1|\widehat N_{12}\widehat N_{22}^j |p_0\kt,
	\label{A-R-series} 
	\end{align}
where 
	\be
	p_1:=k\sin\theta,
	\label{p1-def}
	\ee 
and we have made use of Dirac's bra-ket notation which allows us to express $\check\delta_{p_0}$ and the action of a generic linear operator $\widehat L$ on $\check\delta_{p_0}$ as 
	\begin{align}
	&\check\delta_{p_0}=2\pi \varpi_0|p_0\kt, 
	&&\big(\widehat L\check\delta_{p_0}\big)(p)=2\pi\varpi_0\br p|\widehat L|p_0\kt.
	\label{id-1}
	\end{align}
	
Next, consider the linear operator:
	\be
	\widehat\Pi_k:=\int_{-k}^k dp\:|p\kt\br p|,
	\label{Pi-k-def}
	\ee
which projects $\sF$ onto $\sF_k$ (and $\sF^2$ onto $\sF_k^2$) according to
	\be
	(\widehat\Pi_k\phi)(p)=\left\{\begin{array}{ccc}
	\phi(p)&\for&|p|<k,\\
	0&\for&|p|\geq k.
	\end{array}\right.
	\label{Pi-k}
	\ee
Since, for all $\phi\in\sF_k$,  $\widehat\Pi_k\phi=\phi$, we can identify $\widehat I$ in \eqref{N-ab=} with $\widehat\Pi_k$, and view $\widehat N_{ab}$ as operators acting in $\sF$ that are given by
	\be
	\widehat N_{ab}:=\delta_{ab}\widehat\Pi_k-\widehat M_{ab}.
	\label{N-ab=2}
	\ee
According to \eqref{f=}, \eqref{N-ab=}, and \eqref{B-L-series} -- \eqref{A-R-series}, the fundamental transfer matrix $\widehat\bM$ determines the scattering amplitude $\ff$. 

The central ingredient of DFSS is the curious fact that we can identify $x$ with an evolution parameter (``time'') and express $\widehat\bM$ in terms of the time-evolution operator for an effective non-Hermitian Hamiltonian operator $\widehat\bcH(x)$, \cite{ap-2014,pra-2021,ptep-2024b}. Specifically, {as we show in Ref.~\cite{ptep-2026a},} the fundamental transfer matrix for the scattering problems defined by the Bergmann equation in two dimensions satisfies 
	\be
	\widehat\bM=\widehat\Pi_k\,\widehat\bcU(\infty,-\infty)\widehat\Pi_k,
	\label{dyson-1}
	\ee
where $\widehat\bcU(x,x_0)$ is the evolution operator for the effective Hamiltonian operator, 
	\be
    	\widehat\bcH(x):= \frac{1}{2}e^{-i\widehat\varpi x\bsigma_3}
	\left\{\widehat\sV(x)\widehat\varpi^{-1}\bcK 
	-\widehat\varpi\,\big[\widehat\alpha(x)
	-\widehat I\,\big]\bcK^T\right\}  
	e^{i\widehat\varpi x\bsigma_3},
	\label{bcH-def}
    	\ee
$x$ and $x_0$ respectively play the roles of time and its initial value,	
	\begin{align}
	&\bcK:=i\bsigma_2+\bsigma_3=\left[\begin{array}{cc}
	1 & 1\\
	-1 & -1\end{array}\right],
	&&\bcK^T:=-i\bsigma_2+\bsigma_3=\left[\begin{array}{cc}
	1 & -1\\
	1 & -1\end{array}\right],
	\label{bcK-def}
	\end{align} 
$\bsigma_j$'s denote the Pauli matrices, i.e.,
	\begin{align}
	&\bsigma_1:=\left[\begin{array}{cc}
	0 & 1\\
	1 & 0\end{array}\right],
	&&\bsigma_2:=\left[\begin{array}{cc}
	0 & -i\\
	i & 0\end{array}\right],
	&&\bsigma_3:=\left[\begin{array}{cc}
	1 & 0\\
	0 & -1\end{array}\right],	
	\label{pauli}
	\end{align}
$\widehat\varpi$, $\widehat\sV(x)$, and $\widehat\alpha(x)$ are the linear operators defined by
	\begin{align}
	&\widehat\varpi:=\varpi(\widehat p),
	&&\widehat\sV(x):=\widehat p\big[\widehat{\alpha}(x)^{-1}-\widehat I\,\big] 
	\widehat p-k^2\big[\widehat{\beta}(x)-\widehat I\,\big],
	\label{sV-def}\\
	&\widehat\alpha(x):=\alpha(x, \widehat y),
	&&\widehat\beta(x):=\beta( x,\widehat y),
	\label{alpha-beta-def}
	\end{align}
and $\widehat y$ and $\widehat p$ are respectively the $y$ components of the standard position and momentum operators in two dimensions. These act in $\sF^2$ according to $(\widehat y\,\phi)(p):=i\partial_p\phi(p)$ and $(\widehat p\,\phi)(p):=p\phi(p)$. In particular, $(\widehat\varpi\,\phi)(p)=\varpi(p)\phi(p)$. We also recall that the evolution operator $\widehat\bcU(x,x_0)$ satisfies
	\begin{align}
	&i\partial_x\widehat\bcU(x,x_0)=\widehat\bcH(x)\,\widehat\bcU(x,x_0),
	&&\widehat\bcU(x_0,x_0)=\widehat I.
	\label{sch-eq-U}
	\end{align}  
	
For $x\notin [0,\ell]$, $\alpha(x)=\beta(x)=1$, and in view of \eqref{bcH-def}, \eqref{sV-def}, and \eqref{alpha-beta-def}, we have 
	\be
	\widehat\bcH(x)=\widehat 0~~\for~~x\notin [0,\ell].
	\label{condi-H}
	\ee 
A direct consequence of this observation is that $\widehat\bcU(0,-\infty)=\widehat\bcU(+\infty,\ell)=\widehat I$. This in turn shows that $\widehat\bcU(\infty,-\infty)=\widehat\bcU(\ell,0)$. Substituting this equation in \eqref{dyson-1}, we find	\be
	\widehat\bM=\widehat\Pi_k\,\widehat\bcU(\ell,0)\widehat\Pi_k.
	\label{dyson-1n}
	\ee

Because $x$ plays the role of an effective time variable, $\widehat\bcH(x)$ is a ``time-dependent'' Hamiltonian operator. Therefore, we cannot, in general, find a closed-form expression for its evolution operator. We can only expand it in a Dyson series \cite{sakurai}. Substituting the latter in \eqref{dyson-1n} and making use of \eqref{condi-H}, 
we arrive at {\cite{ptep-2026a}:}
	\be
	\widehat\bM=
	\widehat\Pi_k+\sum_{j=1}^\infty(-i)^j\!\!
	\int_0^\ell \!dx_j\int_0^{x_j}\!\!dx_{j-1}\cdots\int_0^{x_2}\!\!dx_1\:\widehat\Pi_k\,
	\widehat\bcH(x_j)\widehat\bcH(x_{j-1})\cdots\widehat\bcH(x_1)\widehat\Pi_k.
	\label{dyson}
	\ee

\section{Exact solution of the scattering problem for diffraction gratings  \eqref{model-1}}
\label{S3}

\subsection{Explicit form of the effective Hamiltonian $\widehat\bcH(x)$}
\label{S31}

To determine the fundamental transfer matrix \eqref{dyson} for the scattering of TE and TM waves by a grating of the form \eqref{model-1}, we need to study the structure of the effective Hamiltonian $\widehat\bcH(x)$ for $x\in[0,\ell]$. According to  \eqref{bcH-def} and \eqref{sV-def}, this requires the knowledge of the operators $\widehat\alpha(x)$, $\widehat\alpha(x)^{-1}$, and $\widehat\beta(x)$. To find useful expressions for these operators, first we note that, for $x\in[0,\ell]$,
	\begin{align}
	&\alpha(x,y)=\alpha_0+\sum_{n=1}^N \alpha_n(x)e^{in\fK\,y},
	&&\beta(x,y)=\beta_0+\sum_{n=1}^N \beta_n(x)e^{in\fK\,y},
	\label{alpha-beta-sum}
	\end{align}
where, for all $n\geq 0$,
	\begin{align}
	&\alpha_n:=\delta_{0n}+\left\{\begin{array}{cc}
	\fb_n&\mbox{for TE waves},\\
	\fa_n&\mbox{for TM waves},
	\end{array}\right.
	&&\beta_n:=\delta_{0n}+\left\{\begin{array}{cc}
	\fa_n&\mbox{for TE waves},\\
	\fb_n&\mbox{for TM waves},
	\end{array}\right.
	\label{alpha-beta-n=}
	\end{align}
and we have made use of \eqref{model-1} and \eqref{alpha-beta}.\footnote{Note that $\alpha_0$ and $\beta_0$ are constants, while for $n\geq 1$, $\alpha_n$ and $\beta_n$ are functions of $x$.}
	
Next, suppose that, for all $x\in[0,\ell]$, 
	\begin{align}
	\sum_{n=1}^N|\alpha_n(x)|<|\alpha_0|,
	\label{condi}
	\end{align} 
which in particular entails $\alpha_0\neq 0$. Then, we can use the first equation in \eqref{alpha-beta-sum} to arrive at the following series expansion of $\alpha(x,y)^{-1}$.
	\begin{align}
	\alpha(x,y)^{-1} &=\sum_{s=0}^\infty 
	\frac{[-g(x,e^{i\fK y})]^s}{\alpha_0^{s+1}}
	%\nn\\&
	=\frac{1}{\alpha_0}+\sum_{m=1}^\infty \gamma_m(x)e^{im\fK\,y},
	\label{alpha-inv}
	\end{align}
where 
	\begin{align}
	&g(x,z):= \sum_{n=1}^N\alpha_n(x)z^n, 
	&&\gamma_m(x):=\frac{1}{m!}\,\partial_z^m\Gamma(x,z)\Big|_{z=0},
	&&\Gamma(x,z):=\sum_{s=1}^\infty
	\frac{(-1)^s g(x,z)^s}{\alpha_0^{s+1}}.
	\label{eq38}
	\end{align}	
With the help of the identities,
	\begin{align}
	&\partial_z^m[g(x,z)^s]=	
	\sum_{\substack{k_1+\cdots+k_s = m \\ k_i \geq 0}}
\frac{m!}{k_1! k_2! \cdots k_s!}
\, \left[\partial_z^{k_1}g(x,z)\right]\left[\partial_z^{k_2}g(x,z)\right]\cdots 
\left[\partial_z^{k_s}g(x,z)\right],\nn\\
	&\partial_z^kg(x,0)=\left\{\begin{array}{cc}
	k!\,\alpha_k(x)&\for~1\leq k\leq N,\\
	0&{\rm otherwise},\end{array}\right.\nn
	\end{align}
and \eqref{eq38}, we can express $\gamma_m$ in the form
	\begin{align}
	&\gamma_m(x)=
	\sum_{s=1}^m\Bigg[\frac{(-1)^s}{\alpha_0^{s+1}}\hspace{-6pt}
	\sum_{\substack{k_1+\cdots+k_s = m \\ k_i \geq 1}}\hspace{-16pt}
	\alpha_{k_1}(x)\alpha_{k_2}(x)\cdots\alpha_{k_s}(x)\Bigg].
	\label{gamma=}
	\end{align}	
	
In view of \eqref{alpha-beta-def}, \eqref{alpha-beta-sum}, and \eqref{alpha-inv},
	\begin{align}
	&\widehat\alpha(x)=\alpha_0 \widehat I+
	\sum_{n=1}^N\alpha_n(x)\widehat S^n ,
	\label{alpha-op}\\
	&\widehat\beta(x)=\beta_0 \widehat I+
	\sum_{n=1}^N\beta_n(x)\widehat S^n ,
	\label{beta-op}\\
	&\widehat\alpha(x)^{-1} =\alpha_0^{-1}\widehat I+
	\sum_{m=1}^\infty \gamma_m(x)\widehat S^m,
	\label{alpha-inv-op}
	\end{align}
where $\widehat S$ is defined by
	\be
	\widehat S:=e^{i\fK\,\widehat y}.
	\label{S-def}
	\ee
It is easy to see that this is a translation operator; for all $\phi\in\sF^2$ and $p\in\R$,  
	\be
	(\widehat S\phi)(p)=\phi(p-\fK).
	\label{shift}
	\ee
By virtue of \eqref{Pi-k} and \eqref{shift}, it satisfies
	\begin{align}
	&\widehat\Pi_k\widehat S^m\widehat\Pi_k=\widehat 0~~\for~~m\geq \frac{2k}{\fK}.
	\label{terminate}
	\end{align}
Furthermore, given a function $g:\R^2\to\C$, we can use \eqref{Pi-k} and \eqref{shift} to show that 
	\begin{align}
	&\widehat \Pi_k g(x,\widehat p\,)=g(x,\widehat p\,)\widehat\Pi_k,
	&&\widehat S g(x,\widehat p\,)=g(x,\widehat p-\fK\,\widehat I\,)\widehat S.
	\label{S-commutator}
	\end{align}
	
Substituting \eqref{alpha-op} -- \eqref{alpha-inv-op} in \eqref{bcH-def} and \eqref{sV-def} and making use of \eqref{S-commutator}, we can express the effective Hamiltonian $\widehat\bcH(x)$ in the form
	\be
	\widehat\bcH(x)=\bcH_0(x,\widehat p\,)+\sum_{m=1}^\infty \bcH_m(x,\widehat p\,)\widehat S^m,
	\label{H-expand}
	\ee
where, for all $m\in\N:=\{0,1,2,\cdots\}$,
	\begin{align}
	&\bcH_m(x,p):=\chi_\ell(x)\,e^{-ix\varpi(p)\bsigma_3}\bH_m(x,p)
	e^{ix\varpi_m(p)\bsigma_3},
	\label{bcH-m-def}\\[6pt]
	&\bH_m(x,p):=\frac{1}{2}\times\left\{\begin{array}{ccc}
	h_0(p)\varpi(p)^{-1}\bcK-(\alpha_0-1)\varpi(p)\,\bcK^T&\for& m=0,\\[3pt]
	h_m(x,p)\varpi_m(p)^{-1}\bcK-\alpha_m(x)\varpi(p)\bcK^T &\for& 1\leq m\leq N,\\[3pt]
	h_m(x,p)\varpi_m(p)^{-1}\bcK &\for& m\geq N+1,\end{array}\right.
	\label{bHm-def}\\[6pt]
	&h_m(x,p):=\left\{\begin{array}{ccc}
	(\alpha_0^{-1}-1)p^2-(\beta_0-1)k^2&\for&m=0,\\[3pt]
	\gamma_m(x)p(p-m\fK)-k^2\beta_m(x)&\for&1\leq m\leq N,\\[3pt]
	\gamma_m(x)p(p-m\fK)&\for& m\geq N+1,\end{array}\right.
	\label{h-m-def}\\[6pt]
	&\varpi_m(p):=\varpi(p-m\fK)=\left\{\begin{array}{ccc}
	\sqrt{k^2-(p-m\fK)^2}&\for& |p-m\fK|<k,\\
	i\sqrt{(p-m\fK)^2-k^2}&\for&|p-m\fK|\geq k,\end{array}\right.
	\label{varpi-m}
	\end{align}
and $\chi_\ell$ is the function given by \eqref{chi}.

\subsection{Fundamental transfer matrix for a homogeneous slab}
\label{S32}
	
For the special cases where $\alpha_n=\beta_n=0$ for $n\geq 1$, $\alpha$, $\beta$, and consequently the permittivity and permeability of the slab under consideration do not depend on $x$, i.e., it is a homogeneous slab. This implies that its scattering problems for TE and TM waves can be reduced to the corresponding one-dimensional scattering problems which are studied in Ref.~\cite{ptep-2024b}. Furthermore, in this case $\gamma_n=h_n=0$, $\bH_n=\bcH_n=\bzero$, where $\bzero$ stands for the zero matrix, and \eqref{H-expand} reduces to $\widehat\bcH(x)=\bcH_0(x,\widehat p)$. We can obtain an explicit expression for $\bcH_0(x,p)$ by substituting \eqref{bcK-def} in \eqref{bHm-def}, setting $m=0$, and making use of \eqref{bcH-m-def}, \eqref{h-m-def}, and \eqref{varpi-m}. The result is
	\begin{align}
	\bcH_0(x,p)&=e^{-ix\varpi(p)\bsigma_3}\,\bcH_0(0,p)\,e^{ix\varpi(p)\bsigma_3}
	\label{bcH-zero=1}\\
	&=\chi_\ell(x)\varpi(p)\left[\begin{array}{cc}
	- \fm_{+}(p)+1&- \fm_{-}(p)\,e^{-2i x\varpi(p) }\\  \fm_{-}(p)\,e^{2ix \varpi(p)}& \fm_{+}(p)-1\end{array}\right],
	\label{bcH-zero=}
	\end{align}
where
	\begin{align}
	\fm_{\pm}(p)&:= \frac{\tilde{\fn}(p)^2\pm\alpha_0^2}{2\alpha_0}
	=\frac{1}{2\alpha_0}\left(1\pm\alpha_0^2+
	\frac{\alpha_0\beta_0-1}{1-p^2/k^2}\right),
	\label{m-pm-def}\\
	\tilde{\fn}(p)&:=\pm\,\sqrt{\frac{k^2(\alpha_0\beta_0-1)}{\varpi(p)^{2}}+1}
	=\pm\,\sqrt{1+\frac{\alpha_0\beta_0-1}{1-p^2/k^2}}.
	\label{tn-def-0}
	\end{align}
The sign in the latter equation is to be chosen so that the real part of $\tilde\fn(p)$ has the same sign as that of the {complex refractive index of the homogeneous slab which equals $\alpha_0\beta_0$, \cite{ptep-2024b}.\footnote{The negative sign corresponds to a slab made of a negative-index metamaterial \cite{veselago,smith-2008,ap-2016}.}}

For $p=0$, \eqref{bcH-zero=} reduces to Eq.~(30) of Ref.~\cite{ptep-2024b} which gives the effective Hamiltonian used in the dynamical formulation of the scattering of TE and TM waves by an effectively one-dimensional material. This suggests that we can pursue a similar approach to compute the evolution operator $\widehat\bcU_0(x,x_0)$ for the Hamiltonian operator $\bcH_0(x,\widehat p\,)$; using \eqref{bcH-zero=}, we can easily check that
	\be
	\widehat\bcU_0(x,x_0):=e^{-ix\widehat\varpi\bsigma_3}e^{-i(x-x_0)\bcH'_0(\widehat p)}e^{ix_0\widehat\varpi\bsigma_3}
	\label{U-zero=}
	\ee
satisfies 
	\begin{align}
	&i\partial_x\,\widehat\bcU_0(x,x_0)=\bcH_0(x,\widehat p)\,\widehat\bcU_0(x,x_0),
	&&\widehat\bcU_0(x_0,x_0)=\widehat I,
	\label{sch-eq-U-zero}
	\end{align}
where
	\begin{align}
	\bcH'_0(p)&:=\chi_\ell(x)\big[\bcH_0(0,p)-\varpi(p)\bsigma_3\big]
	%\red{=
	%\chi_\ell(x)\varpi(p)\left[\begin{array}{cc}
	%- \fm_{+}(p)&- \fm_{-}(p) \\  \fm_{-}(p) & \fm_{+}(p)\end{array}\right]}\nn\\
	%&
	=\chi_\ell(x)\varpi(p)[i\fm_-(p)\bsigma_2-\fm_+(p)\bsigma_3].\nn
	\end{align}
Substituting this relation in \eqref{U-zero=}, we find
	\be
	\widehat\bcU_0(x,0)=e^{-ix\widehat\varpi\bsigma_3}e^{-ix\bcH_0'(\widehat p)}=
	\bcU_0(x,\widehat p)~~~\for~~~x\in[0,\ell],
	\label{bU0=1}
	\ee
where
	\begin{align}
	\bcU_0(x,p)&:
	%\red{=e^{-ix\varpi(p)\bsigma_3}\left[\!\!\begin{array}{cc}
	%\fc(x,p)+i\fm_{+}(p)\fs(x,p)& i \fm_{-}(p)\fs(x,p)\\[3pt]
	%- i \fm_{-}(p)\fs(x,p)& \fc(x,p)-i\fm_{+}(p)\fs(x,p)
	%\end{array}\!\!\right]}\nn \\&
	=e^{-ix\varpi(p)\bsigma_3}\Big\{
	\fc(x,p)\bI+\fs(x,p)\big[-\fm_-(p)\bsigma_2+i\fm_+(p)\bsigma_3\big]\Big\},
	\label{bU0-solution-2}\\
	\fc(x,p)&:=\cos\!\big[\varpi(p)\tilde\fn(p)x\big],\qquad\qquad
	\fs(x,p):=\frac{\sin\!\big[\varpi(p)\tilde\fn(p)x\big]}{\tilde\fn(p)},
	\label{bU0-solution-2n}
	\end{align}
and $\bI$ stands for the $2\times 2$ identity matrix.
	
In view of \eqref{dyson-1n} and \eqref{bU0=1}, the fundamental transfer matrix describing the scattering of TE and TM waves by the homogeneous slab has the form
	\begin{align}
	\widehat\bM_0=\bM_0(\widehat p)\,\widehat\Pi_k,
	\label{M-zero}
	\end{align}
where
	\be
	\bM_0(p):=\bcU_0(\ell,p),
	\label{M-zero-def}
	\ee
and we have also made use of the fact that $\widehat\Pi_k$ commutes with any function of $\widehat p$ and $\widehat\Pi_k^2=\widehat\Pi_k$.

\subsection{Fundamental transfer matrix for gratings given by \eqref{model-1}}
\label{S33}
	
Consider using \eqref{H-expand} to compute the terms of the Dyson series expansion \eqref{dyson} of the fundamental transfer matrix. Then, by virtue of \eqref{terminate} and \eqref{S-commutator}, we find
	\be
	\int_0^\ell \!dx_j\int_0^{x_j}\!\!dx_{j-1}\cdots\int_0^{x_2}\!\!dx_1\:\widehat\Pi_k\,
	\widehat\bcH(x_j)\widehat\bcH(x_{j-1})\cdots\widehat\bcH(x_1)\widehat\Pi_k=
	\sum_{n=0}^{\lfloor\!\lfloor 2k/\fK \rfloor\!\rfloor} \bG_{jn}(\widehat p)
	\,\widehat\Pi_k\widehat S^{n}\widehat\Pi_k,
	\label{HHH=}
	\ee
where {$\lfloor\!\lfloor x \rfloor\!\rfloor$ stands for the largest integer that is strictly smaller than} $x$, and $\bG_{jn}$ are $2\times 2$ matrix-valued functions of $p$.{\footnote{{$\lfloor\!\lfloor x \rfloor\!\rfloor:=\lceil x \rceil-1$, where $\lceil x \rceil$ is the smallest integer that is not smaller than $x$.}}} It is not difficult to see from \eqref{H-expand} and  \eqref{HHH=} that
	\be
	\bG_{j0}(p)=\int_0^\ell \!dx_j\int_0^{x_j}\!\!dx_{j-1}\cdots\int_0^{x_2}\!\!dx_1\: 
	\bcH_0(x_j, p)\bcH_0(x_{j-1}, p)\cdots \bcH_0(x_1, p).
	\label{Gn0}
	\ee
For $k<\fK/2$ and $n\geq 1$, $\widehat\Pi_k\widehat S^{n}\widehat\Pi_k=\widehat 0$, 
the right-hand side of \eqref{HHH=} coincides with $\bG_{j0}(\widehat p)$, and \eqref{dyson}, \eqref{HHH=}, and \eqref{Gn0} give $\widehat\bM=\widehat\bM_0$. Consequently, the incident TE and TM waves are not affected by the inhomogeneities of the grating. In other words, the scattering amplitude for $k<\fK/2$ is identical to the one for the homogeneous slab of the preceding subsection.

To determine the fundamental transfer matrix for $k\geq\fK/2$, we substitute \eqref{HHH=} in \eqref{dyson}. This gives
	\be
	\widehat\bM=\widehat\bM_0+
	\sum_{n=1}^{\lfloor\!\lfloor 2k/\fK \rfloor\!\rfloor}\bM_n(\widehat p\,)\,\widehat\Pi_k\widehat S^n\widehat\Pi_k,
	\label{M-truncate}
	\ee
where %$\widehat\bM_0$ is given by \eqref{M-zero}, and %for $1\leq n\leq \lfloor\!\lfloor 2k/\fK \rfloor\!\rfloor$,
	\be
	\bM_n(p):=\sum_{j=1}^\infty(-i)^j\bG_{jn}(p).
	\label{Mn-def-1}
	\ee
	
Next, consider the special cases where $\alpha_0=\beta_0=1$. Then \eqref{bcH-zero=} and \eqref{m-pm-def} imply $\bcH_0(x,p)=\bzero$, and we can use \eqref{H-expand} and \eqref{U-zero=} --  
\eqref{M-zero-def} to show that
	\be
	\widehat\bcH(x)=\sum_{m=1}^\infty \bcH_m(x,\widehat p\,)\widehat S^m,
	\label{H-expand-zero}
	\ee
$\bcU_0(x,\widehat p)=\widehat I$, and
	\be
	\widehat\bM_0=\widehat\Pi_k.
	\label{M-zero-zero}
	\ee
 According to \eqref{terminate}, \eqref{S-commutator}, and \eqref{H-expand-zero}, the left-hand side of \eqref{HHH=} vanishes for $j>\lfloor\!\lfloor 2k/\fK \rfloor\!\rfloor$. This means that $\bG_{jn}=0$ for $j>\lfloor\!\lfloor 2k/\fK \rfloor\!\rfloor$, and \eqref{Mn-def-1} gives
	\be
	\bM_n(p)=\sum_{j=1}^{\lfloor\!\lfloor 2k/\fK \rfloor\!\rfloor}(-i)^j\bG_{jn}(p).
	\label{Mn=}
	\ee
	
Equations \eqref{M-truncate}, \eqref{M-zero-zero}, and \eqref{Mn=} reduce the determination of the fundamental transfer matrix for $\alpha_0=\beta_0=1$ to the calculation of the matrix-valued functions $\bG_{jn}$ with $j,n\in\{1,2,\cdots \lfloor\!\lfloor 2k/\fK \rfloor\!\rfloor\}$. This involves evaluating the integrals in \eqref{HHH=} for $j\in\{1,2,\cdots, \lfloor\!\lfloor 2k/\fK \rfloor\!\rfloor\}$.
	
This analysis does not apply to the general case, where $\alpha_0\neq 1$ or $\beta_0\neq 1$, which is of greater practical significance.\footnote{$\alpha_0=\beta_0=0$ corresponds to active gratings involving both regions of gain and loss which would be practically difficult to engineer.} Fortunately, there is a transformation scheme for mapping the general case to a special case which shares the features of {the cases with} $\alpha_0=\beta_0=1$ that are responsible for its exact solvability. In the remainder of this subsection, we discuss the details of this scheme.
		
Consider the following transformation of the effective Hamiltonian $\widehat\bcH(x)$.
	\begin{align}
	&\widehat\bcH(x)\to\widehat{\bsH}(x)=\widehat\bcU_0(x,x_0)^{-1}
	\big[\widehat\bcH(x)-\bcH_0(x,\widehat p)\big]\,\widehat\bcU_0(x,x_0),
	\label{H-trans}
	\end{align}
where $\widehat\bcU_0(x,x_0)$ is the evolution operator for $\bcH_0(x,\widehat p)$. Then, by virtue of \eqref{sch-eq-U} and \eqref{sch-eq-U-zero}, the evolution operator $\widehat{\bsU}(x,x_0)$ for the transformed Hamiltonian $\widehat{\bsH}(x)$ takes the form
	\begin{align}
	\widehat{\bsU}(x,x_0)=
	\widehat\bcU_0(x,x_0)^{-1}\widehat\bcU(x,x_0).
	\label{U-trans}
	\end{align}
	
Setting $x_0=0$ in \eqref{H-trans} and \eqref{U-trans} and making use of 
\eqref{S-commutator}, \eqref{H-expand}, and \eqref{bU0=1},
%\eqref{dyson-1n},  \eqref{S-commutator}, \eqref{H-expand}, \eqref{M-zero} and \eqref{M-zero-def}, 
we obtain 
	\begin{align}
	&\widehat{\bsH}(x)=\sum_{m=1}^\infty {\bsH_m}(x,\widehat p)\widehat S^m,
	\label{H-trans-2}\\
	&\widehat\bcU(x,0)= \bcU_0(x,\widehat p)\,\widehat{\bsU}(x,0),
	\label{U-trans-2}
	\end{align}
 where
 	\begin{align}
	&{\bsH_m}(x,p):=\bcU_0(x,p)^{-1}{\bcH_m}(x,p)\,\bcU_0(x,p-m\fK).
	\label{sH-m=}
	\end{align}
Equations \eqref{dyson-1n}, \eqref{M-zero}, and \eqref{U-trans-2} lead to the following expression for the fundamental transfer matrix.
	\be
	\widehat\bM=\widehat\bM_0\,\widehat{\bsM},
	\label{M=MM}
	\ee
where
	\begin{align}
	\widehat{\bsM}&:=\widehat\Pi_k\,\widehat{\bsU}(\ell,0)\,\widehat\Pi_k\nn\\
	&=\widehat\Pi_k+\sum_{j=1}^\infty(-i)^j\!\!
	\int_0^\ell \!dx_j\int_0^{x_j}\!\!dx_{j-1}\cdots\int_0^{x_2}\!\!dx_1\:\widehat\Pi_k\,
	\widehat{\bsH}(x_j)\widehat{\bsH}(x_{j-1})\cdots\widehat{\bsH}(x_1)\widehat\Pi_k.
	\label{new-M}
	\end{align}
 
Next, we recall the argument we used to show that, for $\alpha_0=\beta_0=1$, the Dyson series \eqref{dyson} for fundamental transfer matrix truncates and that we can compute its terms in an exact manner. This argument rests on the premise that the smallest power of the operator $\widehat S$ that contributes to the series expansion \eqref{H-expand-zero} of the effective Hamiltonian $\widehat{\bcH}(x)$ is greater than or equal to $1$. According to \eqref{H-trans-2}, this is also true for the transformed Hamiltonian $\widehat{\bsH}(x)$. Therefore, we can similarly argue that the right-hand side of \eqref{new-M} truncates, and we obtain the following exact expression for $\widehat{\bsM}$.
	\be
	\widehat\bsM=\widehat\Pi_k+
	\sum_{n=1}^{\lfloor\!\lfloor 2k/\fK \rfloor\!\rfloor}\bsM_n(\widehat p\,)\,\widehat\Pi_k\widehat S^n\widehat\Pi_k,
	\label{sM-exact}
	\ee
where $\bsM_n$ are $2\times 2$ matrix-valued functions of $p$ that are determined by the first $\lfloor\!\lfloor 2k/\fK \rfloor\!\rfloor$ terms in the sum on the right-hand side of \eqref{new-M}. Substituting \eqref{M-zero} and \eqref{sM-exact} in \eqref{M=MM},  we arrive at an exact expression for $\widehat\bM$ that consists of a finite sum of operators.

\subsection{Exact solvability of the gratings given by \eqref{model-1}}
\label{S34}

To obtain the exact solution of the scattering problem, we first outline the calculation of the scattering amplitude for the special cases where $\alpha_0=\beta_0=1$, where we can compute the fundamental transfer matrix using \eqref{M-truncate},
\eqref{M-zero-zero}, and \eqref{Mn=}. To this end, we examine the structure of the linear operators $\widehat N_{ab}$ defined by \eqref{N-ab=2}. Substituting \eqref{M-truncate}  in this relation and making use of \eqref{M-zero-zero}, we find
	\be
	\widehat N_{ab}=\sum_{n=1}^{\lfloor\!\lfloor 2k/\fK \rfloor\!\rfloor}N_{n,ab}(\widehat p\,)\,\widehat\Pi_k\widehat S^n\widehat\Pi_k,
	\label{Nab-101}
	\ee
where $N_{n,ab}(p)$ are the entries of the matrices $-\bM_n(p)$.

In Appendix~A, we prove that, for every pair of positive integers $m$ and $n$, and generic functions $u:\R\to\C$, the following identity holds.
	\be
	\widehat\Pi_k\widehat S^m u(\widehat p)\,\widehat\Pi_k \widehat S^n \widehat\Pi_k=
	u(\widehat p-m\fK\,\widehat I\,)\,\widehat\Pi_k\,\widehat S^{m+n}\widehat\Pi_k.	
	\label{id-101}
	\ee
According to \eqref{terminate}, \eqref{Nab-101}, and \eqref{id-101},
	\be
	\widehat N_{a_1b_1}\widehat N_{a_2b_2}\cdots \widehat N_{a_jb_j}=\widehat 0~~\for~~ j>\lfloor\!\lfloor 2k/\fK \rfloor\!\rfloor.
	\label{condi-N}
	\ee 
Consequently, the series expansions \eqref{B-L-series} --  \eqref{A-R-series} of the coefficient functions $A^{l/r}_+$ and $B^{l/r}_-$ truncate, and we can obtain exact expressions for them. Substituting these in \eqref{f=}, we arrive at an exact expression for the scattering amplitude for $\alpha_0=\beta_0=1$.

This argument does not apply to more general cases in which $\alpha_0$ or $\beta_0$ may differ from 1. For these cases, the fundamental transfer matrix takes the form \eqref{M=MM}. In view of \eqref{N-ab=2} and \eqref{sM-exact}, this implies 
	\be
	\widehat N_{ab}=\delta_{ab}\widehat\Pi_k-M_{0,ab}(\widehat p\,)
	-\sum_{n=1}^{\lfloor\!\lfloor 2k/\fK \rfloor\!\rfloor}\sum_{c=1}^2
	M_{0,ac}(\widehat p\,)\,\sM_{n,cb}(\widehat p\,)\,\widehat\Pi_k\widehat S^n\widehat\Pi_k,
	\label{N-gen=}
	\ee
where $M_{0,ab}(p)$ and $\sM_{n,ab}(p)$ are respectively the entries of $\bM_0(p)$ and $\bsM_n(p)$. Because $\bM_0(\widehat p)\neq\widehat I$, \eqref{N-gen=} does not imply \eqref{condi-N}, and the series expansions \eqref{B-L-series} and  \eqref{A-R-series} for $A^{l/r}_+$ and $B^{l/r}_-$ do not truncate. There is, however, an alternative method for computing these coefficients which we describe in the sequel.

According to \eqref{As=} and \eqref{Bs=}, the determination of $A^{l/r}_+$ and $B^{l/r}_-$ requires the computation of $\widehat M_{22}^{-1}$. {This is} because we can write the solutions of \eqref{Bs=} as
	\begin{align}
	&B^l_-=-\widehat M_{22}^{-1}\widehat M_{21}\check\delta_{p_0}.
	&B^r_-=\widehat M_{22}^{-1}\check\delta_{p_0}.
	\label{B-right}
	\end{align}
	
To determine $\widehat M_{22}^{-1}$, we introduce the linear operators $\widehat\sN_{ab}:\sF_k\to\sF_k$ that are given by
	\be
	\widehat\sN_{ab}:=\delta_{ab}\widehat I-\widehat\sM_{ab}.
	\label{sN-ab}
	\ee
We can extend the definition of these operators to $\sF$ by identifying them with
	\begin{align}
	\widehat\sN_{ab}&:=\delta_{ab}\widehat\Pi_k-\widehat\sM_{ab}
	=-\sum_{n=1}^{\lfloor\!\lfloor 2k/\fK \rfloor\!\rfloor}\bsM_{n,ab}(\widehat p\,)\,\widehat\Pi_k\widehat S^n\widehat\Pi_k.
	\label{sN-ab-2}
	\end{align}	
In view of \eqref{id-101} and \eqref{sN-ab-2}, 
	\be
	\widehat\sN_{a_1b_1}\widehat \sN_{a_2b_2}\cdots \widehat \sN_{a_jb_j}=\widehat 0~~\for~~ j>\lfloor\!\lfloor 2k/\fK \rfloor\!\rfloor.
	\label{condi-sN}
	\ee 
We can use this identity together with \eqref{tdelta}, \eqref{m-pm-def}, \eqref{bU0-solution-2} -- \eqref{M-zero-def}, \eqref{M=MM}, and \eqref{sN-ab} to show that
	\begin{align}
	\widehat M_{22}&=\widehat M_{0,22}\big(\widehat\sM_{22}+
	\widehat M_{0,22}^{-1}\widehat M_{0,21}\widehat\sM_{12}\big)
	%\nn\\&
	={T}(\widehat p)^{-1}\big[\widehat I-\widehat\sN_{22}
	+R^l(\widehat p)\widehat\sN_{12}\big],
	\label{M22=2}\\
	\widehat M_{22}^{-1}&=\big[\widehat I-\widehat\sN_{22}
	+R^l(\widehat p)\widehat\sN_{12}\big]^{-1}{T}(\widehat p)
	%\nn\\&
	=\!\!\sum_{j=0}^{\lfloor\!\lfloor 2k/\fK \rfloor\!\rfloor}\!\!  \big[\widehat\sN_{22}-R^l(\widehat p)\widehat\sN_{12}\big]^j {T}(\widehat p),
	\label{M22=2-inv}
	\end{align}
where $\widehat M_{0,ab}:=M_{0,ab}(\widehat p)$ and 
	\begin{align}
	&{T}(p):=\frac{1}{M_{0,22}(p)}=\frac{\tilde\fn(p)\,e^{-i\ell\varpi(p)}}{
	\tilde\fn(p)\cos[\ell\varpi(p)\tilde\fn(p)]-i
	\fm_+(p)\sin[\ell\varpi(p)\tilde\fn(p)]},
	\label{T-def}\\
	&R^l(p):=-\frac{M_{0,21}(p)}{M_{0,22}(p)}=
	%\frac{i\fm_-(p)\fs(\ell ,p)}{\fc(\ell ,p)-i\fm_+(p)\fs(\ell ,p)}=
	\frac{i\fm_-(p)}{\tilde\fn(p)\cot[\ell\varpi(p)\tilde\fn(p)]-i\fm_+(p)}.
	\label{R-L-def}
	\end{align}
	
Let us also introduce
	\begin{align}
	&R^r(p):=\frac{M_{0,12}(p)}{M_{0,22}(p)}=
	%\frac{i\fm_-(p)\fs(\ell ,p)}{\fc(\ell ,p)-i\fm_+(p)\fs(\ell ,p)}=
	\frac{i e^{-2i\ell\varpi(p)}\fm_-(p)}{\tilde\fn(p)\cot[\ell\varpi(p)\tilde\fn(p)]-i\fm_+(p)},
	\label{R-R-def}\\
	&\sT(\varphi):=T(k\sin\varphi),
	\quad\quad \sR^{l/r}(\varphi):=R^{l/r}(k\sin\varphi).
	\label{Rs-T-zero}
	\end{align}
Then $\sT(\theta_0)$ and $\sR^{l/r}(\theta_0)$ are respectively the transmission and left/right reflection amplitudes of the homogeneous slab corresponding to $\alpha_n=\beta_n=0$ for $n\geq 1$, \cite{jpa-2025}. 

In Appendix~B, we {obtain analogous expressions to \eqref{M22=2} for $\widehat M_{11}$,  $\widehat M_{12}$, and  $\widehat M_{22}$, and use them together with \eqref{As=}, \eqref{B-right}, and \eqref{M22=2-inv} to} derive the following formulas for $A_+^{l/r}$ and $B_-^{l/r}$.
	\begin{align}
	&\begin{aligned}
	A_+^l(k\sin\theta)&=2\pi\big[\sT(\theta_0)\delta(\theta-\theta_0)
	-\sT(\theta)\big\{\cN_{11}(\theta)
	+\sR^l(\theta_0)[\cN_{12}(\theta)+\cV(\theta)]+\cW(\theta)\big\}\big],
	\end{aligned}
	\label{A-left-2}\\
	&\begin{aligned}
	B_-^l(k\sin\theta)=\:&
	2\pi\big[\sR^l(\theta_0)\delta(\theta-\theta_0)+
	\sR^l(\theta_0)\cX(\theta)+\cY(\theta)\big],
	\end{aligned}
	\label{B-left-2}\\
	&\begin{aligned}
	A_+^r(k\sin\theta)&=2\pi\big[\sR^r(\theta_0)\,\delta(\theta-\theta_0)
	-\sT(\theta_0)\sT(\theta)\big\{\cN_{12}(\theta)+\cV(\theta)\big\}\big],
	\end{aligned}
	\label{A-right-2}\\
	&\begin{aligned}
	B_-^r(k\sin\theta)&=
	2\pi \sT(\theta_0)\big[\delta(\theta-\theta_0)+\cX(\theta)\big],
	\end{aligned} 
	\label{B-right-2}
	\end{align}
where
	\begin{align}
	%&\sS(\theta):=\sT(\theta)M_{0,11}(k\sin\theta),
	%\label{sS-def}\\	
	&\cN_{ab}(\theta):=k|\cos\theta_0|\,\br p_1|\widehat\sN_{ab}|p_0\kt,
	\label{cN-def}\\[6pt]
	&\cV(\theta):=k|\cos\theta_0|\sum_{j=2}^{\lfloor\!\lfloor 2k/\fK \rfloor\!\rfloor}\!\!
	\br p_1|\widehat\sN_{12}
	\big[\widehat\sN_{22}-R^l(\widehat p)\widehat\sN_{12}\big]^{j-1} |p_0\kt,
	\label{cV-def}\\
	&\cW(\theta):=k|\cos\theta_0|
	\sum_{j=2}^{\lfloor\!\lfloor 2k/\fK \rfloor\!\rfloor}\!\!\br p_1|\widehat\sN_{12}
	\big[\widehat\sN_{22}-R^l(\widehat p)\widehat\sN_{12}\big]^{j-2}
	\big[\widehat\sN_{21}-R^l(\widehat p)\widehat\sN_{11}\big]|p_0\kt,
	\label{cW-def}\\
	&\cX(\theta):=k|\cos\theta_0|\sum_{j=1}^{\lfloor\!\lfloor 2k/\fK \rfloor\!\rfloor}\!\!
	\br p_1|\big[\widehat\sN_{22}-R^l(\widehat p)\widehat\sN_{12}\big]^j |p_0\kt,\\
	&\cY(\theta):=k|\cos\theta_0|\sum_{j=1}^{\lfloor\!\lfloor 2k/\fK \rfloor\!\rfloor}\!\!
	\br p_1|\big[\widehat\sN_{22}-R^l(\widehat p)\widehat\sN_{12}\big]^{j-1} 
	\big[\widehat\sN_{21}-R^l(\widehat p)\widehat\sN_{11}\big]
	|p_0\kt.
	\label{cZ-def}
	\end{align}
	
Next, let $f_j:\R\to\C$ be a function for each $j\in\Z^+$. Then, in view of  \eqref{tdelta}, \eqref{p1-def}, \eqref{Pi-k-def}, \eqref{shift}, and the identity
	\be
	\delta(p_1-p_0)=\delta\big(k(\sin\theta-\sin\theta_0)\big)=
	\frac{\delta(\theta-\theta_0)+\delta(\theta+\theta_0-180^\circ)}{k|\cos\theta_0|},
	\label{delta-id}
	\ee
we have 
	\begin{align}
	\sum_{j=1}^{\lfloor\!\lfloor 2k/\fK \rfloor\!\rfloor} \br p_1| f_j(\widehat p)\,\widehat\Pi_k\widehat S^j
	\widehat\Pi_k|p_0\kt&=
	\sum_{j=1}^{\lfloor\!\lfloor 2k/\fK \rfloor\!\rfloor}  f_j(p_1)\,\br p_1|\widehat S^j|p_0\kt\nn\\
	&=\sum_{j=1}^{\lfloor\!\lfloor 2k/\fK \rfloor\!\rfloor}  f_j(p_1)\,\delta(p_1-j\fK-p_0)\nn\\
	&= \frac{1}{k|\cos\theta_0|}\sum_{j=1}^{J} f_j(k \fs_j)\,
	\big[\delta(\theta-\theta_{j+})+\delta(\theta-\theta_{j-})\big],
	\label{id-324}
	\end{align}
where 
	\begin{align}
	&J:=\lfloor\!\lfloor \tfrac{k}{\fK}(1-\sin\theta_0)\rfloor\!\rfloor,
	&&\theta_{j+}:=\arcsin\fs_j\in(-90^\circ,90^\circ), 
	\label{theta-p-def} \\
	&\fs_j:=\sin\theta_0+\frac{j\fK}{k},
	&&\theta_{j-}:=180^\circ-\theta_{j+}\in(90^\circ,270^\circ).
	\label{sj-def} 
	\end{align}
Furthermore, given a function $f:\R\to\C$ and nonnegative integers $j$ and $m$ satisfying $j\geq m$, we can use \eqref{theta-p-def} and \eqref{sj-def} to show that
	\be
	f(k\sin\theta_{j\pm}-m\fK)=f(k\fs_j-m\fK)=f(k\sin\theta_0+(j-m)\fK)=f(k\fs_{j-m}).
	\label{id-892}
	\ee

Equation~\eqref{id-324} together with \eqref{sN-ab-2},  \eqref{cN-def}, \eqref{delta-id}, and \eqref{id-892} allow us to express $\cN_{ab}$ in the form
	\begin{align}
	\cN_{ab}(\theta)
	%&=-\sum_{j=1}^{J} \sM_{j,ab}(k\sin\theta)
	%\big[\delta(\theta-\theta_{j+})+\delta(\theta-\theta_{j-})\big]\nn\\
	&=-\sum_{j=1}^{J} \sM_{j,ab}(k \fs_{j})
	\big[\delta(\theta-\theta_{j+})+\delta(\theta-\theta_{j-})\big].
	\label{cN=2}
	\end{align}
It is also easy to see from \eqref{cV-def} -- \eqref{cZ-def} that $\cV, \cW, \cX$, and $\cY$ are sums of terms of the form
	\be
	\br p_1| \big[\widehat\sN_{a_1b_1}-\nu_1R^l(\widehat p)\widehat\sN_{c_1d_1}\big]
	\big[\widehat\sN_{a_2b_2}-\nu_2R^l(\widehat p)\widehat\sN_{c_2d_2}\big]
	\cdots
	\big[\widehat\sN_{a_nb_n}-\nu_nR^l(\widehat p)\widehat\sN_{c_nd_n}\big]|p_0\kt,
	\label{generic}
	\ee
where $\nu_1,\nu_2,\cdots,\nu_n\in\{0,1\}$. According to \eqref{id-101} and \eqref{sN-ab-2}, we can express \eqref{generic} as the left-hand side of \eqref{id-324} for some functions $f_n$ of $p$. Therefore, the same holds for $\cV(\theta), \cW(\theta), \cX(\theta), \cY(\theta)$, and consequently $A^{l/r}(k\sin\theta)$ and $B^{l/r}(k\sin\theta)$. We can, therefore, use \eqref{id-324} to conclude that the scattering amplitude for the diffraction gratings given by \eqref{model-1} has the form
	\begin{align}
	&\ff(\theta)=\sum_{j=0}^{J} \big[\tau_{j+} \delta(\theta-\theta_{j+})+
	\tau_{j-} \delta(\theta-\theta_{j-})\big],
	\label{f=2}
	\end{align}
where $\tau_{j\pm}$ are the complex amplitudes of the diffracted beams {which depend on $\theta_0$ and $k$. In Appendix~C, we show that the normalized intensities of the diffracted beams are given by $|\tau_{j\pm}|^2/2\pi$.}

Substituting \eqref{A-left-2} -- \eqref{B-right-2} in \eqref{f=} and making use of \eqref{cV-def} -- \eqref{cZ-def} and \eqref{cN=2}, we can determine $\tau_{j\pm}$. The above argument shows that this only requires performing finitely many algebraic calculations. This completes the proof of the exact solvability of scattering problems for TE and TM waves incident upon a diffraction grating given by \eqref{model-1}. {We end this subsection by commenting on the assumptions and conditions we have employed to arrive at this proof.
	\begin{enumerate}
	\item Our calculation of the scattering amplitude produces a finite result provided that the operator $\widehat M_{22}$ is invertible. According to \eqref{M22=2-inv} -- \eqref{R-L-def}, this is the case provided that $M_{0,22}(p)\neq 0$ for $|p|<k$. This is the condition for the absence of spectral singularities \cite{prl-2009}. The emergence of the latter mark the onset of lasing \cite{jo-2017}. For a grating lacking regions of gain, spectral singularities do not arise and the scattering amplitude (and consequently diffracted beam amplitudes) do not diverge.
	\item Our results apply to every diffraction grating with permittivity and permeability profiles of the form \eqref{model-1} that satisfy \eqref{condi}. This condition, which in view of \eqref{alpha-beta-n=} is equivalent to	
	\be
	\begin{aligned}
	&\sum_{n=1}^N|\fb_n(x)|<|\fb_0+1|~~~\for~~~\mbox{TE waves},\nn\\
	&\sum_{n=1}^N|\fa_n(x)|<|\fa_0+1|~~~\for~~~\mbox{TM waves},\nn
	\end{aligned}
	\label{id-1265}
	\ee
ensures the absolute convergence of the series expansion~\eqref{alpha-inv} of $\alpha^{-1}$. According to \eqref{bcH-def} and \eqref{sV-def}, the latter enters the expressions for the effective Hamiltonian $\widehat\bcH(x)$ and consequently the fundamental transfer matrix and scattering amplitude.  Equations~\eqref{alpha-beta-def}, \eqref{alpha-beta-sum}, and \eqref{alpha-beta-n=} show that for TE waves scattered by a nonmagnetic grating and TM waves scattered by a purely magnetic grating, $\alpha=\alpha_0=1$, $\alpha_n=0$ for all $n\geq 1$, and \eqref{condi} holds automatically. 
	\end{enumerate}}

\subsection{Calculation of the diffracted beam amplitudes}

The quantity $J$ appearing in \eqref{f=2} determines the number of diffracted beams. According to \eqref{theta-p-def} it takes a nonnegative integer value if and only if
	\be
	\sin\theta_0<1-\frac{J\fK}{k}.
	\label{condi-k}
	\ee 
This condition holds trivially for $J=0$. Therefore, the zeroth-order diffracted beams are always present.  The first-order diffracted beams arise provided that \eqref{condi-k} holds for a positive integer $J$. Because $\sin\theta_0>-1$, this can happen only if $k>\fK/2$. Suppose that $k>\fK/2$, then \eqref{condi-k} is equivalent to 
	\be
	-90^\circ<\theta_0<\theta_{J}~~~{\rm or}~~~180^\circ-\theta_{J}<\theta_0<270^\circ,
	\label{condi-94}
	\ee
where 
	\be
	\theta_{J}:=\arcsin\left(1-\tfrac{J\fK}{k}\right)\in[-90^\circ,90^\circ]. 
	\label{theta-J}
	\ee
Therefore, the first-order diffracted beams exist only for the range of values of the incidence angle given by \eqref{condi-94}. In the following, we will assume that this is the case. 

We begin our study of the structure of the diffracted beam amplitudes $\tau_{j\pm}$ by noting that in light of \eqref{M=MM}, \eqref{sN-ab-2}, \eqref{A-left-2} -- \eqref{cZ-def}, and \eqref{f=2}, the inhomogeneities of the grating do not contribute to $\tau_{0\pm}$. This suggests that we can express them in terms of the reflection and transmission amplitudes of the corresponding homogeneous slab. To do this, we use \eqref{f=} and \eqref{A-left-2} -- \eqref{B-right-2}, to write the scattering amplitude of this slab in the form
	\begin{align}
	\ff_0(\theta)=-\sqrt{2\pi}\,i\times\left\{
	\begin{array}{ccc}
	[\sT(\theta_0)-1]\,\delta(\theta-\theta_0)&\for&
	\cos\theta_0\cos\theta>0,\\
	\sR^l(\theta_0)\delta(\theta-\theta_0)&\for&
	\cos\theta_0>0~\&~\cos\theta<0,\\
	\sR^r(\theta_0)\delta(\theta-\theta_0)&\for&
	\cos\theta_0<0~\&~\cos\theta>0.
	\end{array}\right.
	\label{f-zero}
	\end{align}  
Recalling that $\ff_0(\theta)$ equals the sum of the terms on the right-hand side of \eqref{f=2} with $j=0$, we obtain the following formulas for the amplitudes of the zeroth-order diffracted beams.
	\begin{align}
	&\tau_{0+}=-\sqrt{2\pi}\,i\times\left\{\begin{array}{ccc}
	\sT(\theta_0)-1 &\for & \cos\theta_0>0,\\
	\sR^r(\theta_0) &\for & \cos\theta_0<0,
	\end{array}\right.
	\label{tau-0p}\\
	&\tau_{0-}=-\sqrt{2\pi}\,i\times\left\{\begin{array}{ccc}
	\sR^l(\theta_0) &\for & \cos\theta_0>0,\\
	\sT(\theta_0)-1 &\for & \cos\theta_0<0.
	\end{array}\right.
	\label{tau-0m}
	\end{align} 
	
To compute the amplitudes of the higher-order diffracted beams, we substitute \eqref{sN-ab-2} in \eqref{cN-def} -- \eqref{cZ-def} to express the right-hand sides of these equations in the form \eqref{id-324}. This allows us to write \eqref{A-left-2} -- \eqref{B-right-2} in the form
	\begin{align}
	&A_+^{l/r}(k\sin\theta)= \sum_{j=0}^J
	\cA_j^{l/r} \delta(\theta-\theta_{j+})\quad~\for\quad \cos\theta>0,
	\label{A-sum}\\
	&{B_-^{l/r}}(k\sin\theta)= \sum_{j=0}^J
	\cB_j^{l/r} \delta(\theta-\theta_{j-})\quad~\for\quad \cos\theta<0,
	\label{B-sum}
	\end{align}
where $\cA_j^{l/r}$ and $\cB_j^{l/r}$ are certain complex coefficients, and we have benefitted from the fact that $\pm\cos\theta_{j\pm}>0$. Inserting \eqref{A-sum} and \eqref{B-sum} in \eqref{f=} and comparing the resulting equation with \eqref{f=2}, we arrive at
	\begin{align}
	&\tau_{j+}=\frac{-i}{\sqrt{2\pi}}\times\left\{\begin{array}{ccc}
	\cA_j^{l} &\for & \cos\theta_0>0,\\
	\cA_j^{r} &\for & \cos\theta_0<0,
	\end{array}\right.
	\label{tau-jp}\\
	&\tau_{j-}=\frac{-i}{\sqrt{2\pi}}\times\left\{\begin{array}{ccc}
	\cB_j^l &\for & \cos\theta_0>0,\\
	\cB_j^r &\for & \cos\theta_0<0.
	\end{array}\right.
	\label{tau-jm}
	\end{align} 	
It is easy to check that for $j=0$ these relations reproduce \eqref{tau-0p} and \eqref{tau-0m}.

In Appendix~D, we employ the above prescription to compute $\cA^{l/r}_j$ and  $\cB^{l/r}_j$ for $j\in\{1,2\}$. Substituting the resulting equations in \eqref{tau-jp} and \eqref{tau-jm}, we can express  $\tau_{1\pm}$ and $\tau_{2\pm}$ {in terms of $\sM_{j,ab}(k\fs_i)$ with $i,j\in\{1,2\}$}. Here we present the formulas for $\tau_{1\pm}$:
	\begin{align}
	&\tau_{1+}=-\sqrt{2\pi}\,i 
	\,\sT(\theta_{1+})
	\times\left\{\begin{array}{ccc}
	\sM_{1,11}(k\,\fs_1)+\sR^l(\theta_0)\sM_{1,12}(k\,\fs_1)&\for & \cos\theta_0>0,\\
	\sT(\theta_0)	\sM_{1,12}(k\,\fs_1)&\for & \cos\theta_0<0,
	\end{array}\right.
	\label{tau-1p}\\[6pt]
	&\tau_{1-}=\sqrt{2\pi}\,i 
	\times\left\{\begin{array}{ccc}
	\begin{aligned}
	&\sM_{1,21}(k\,\fs_1)+\sR^l(\theta_0)\sM_{1,22}(k\,\fs_1)\\
	&-\sR^l(\theta_{1+})[\sM_{1,11}(k\,\fs_1)+\sR^l(\theta_{0})\sM_{1,12}(k\,\fs_1)]
	\end{aligned} &\for & \cos\theta_0>0,\\[12pt]
	\sT(\theta_{0})[\sM_{1,22}(k\,\fs_1)-\sR^l(\theta_{1+})\sM_{1,12}(k\,\fs_1)] &\for & \cos\theta_0<0.
	\end{array}\right.
	\label{tau-1m}
	\end{align}
Observe that, according to \eqref{Rs-T-zero} and \eqref{id-892}, $\sR^l(\theta_{j\pm})=R^l(k\sin\theta_{j\pm})=R^l(k\fs_j)$. In particular, $\sR^l(\theta_{j+})=
\sR^l(\theta_{j-})$.
	
If $\fK/2<k<\fK$, we can satisfy \eqref{condi-k} only for $J=0$ and $J=1$. Therefore the second- and higher-order diffracted beams are absent. To generate second-order diffracted beams, we must consider incident waves whose wavenumber and incidence angle respectively exceeds $\fK$ and fulfills \eqref{condi-94} for $J=2$.  

{Equations~\eqref{tau-0p}, \eqref{tau-0m}, \eqref{tau-1p}, ad \eqref{tau-1p} reduce the calculation of the amplitudes for the zeroth- and first-order diffracted beams to that of the  functions $\sR^{l/r}$ and $\sT$, which give the reflection and transmission amplitudes of the underlying homogeneous slab, and the entries of the matrix-valued function $\bsM_1$. In the remainder of this subsection, we describe the computation the latter.}
 
First, we use \eqref{bcH-m-def}, \eqref{bU0=1}, \eqref{bU0-solution-2n}, \eqref{H-trans-2}, \eqref{sH-m=}, \eqref{new-M}, and \eqref{sM-exact} to show that
 	\begin{align}
	e^{-ix\bcH_0'(p)}&=e^{ix\varpi(p)\bsigma_3}\bcU_0(x,p)
	%\red{=
	%\left[\!\!\begin{array}{cc}
	%\fc(x,p)+i\fm_{+}(p)\fs(x,p)& i \fm_{-}(p)\fs(x,p)\\[3pt]
	%- i \fm_{-}(p)\fs(x,p)& \fc(x,p)-i\fm_{+}(p)\fs(x,p)
	%\end{array}\!\!\right]}\nn\\
	%&
	=\fc(x,p)\,\bI+\fs(x,p)\left[-\fm_{-}(p)\,\bsigma_2+i\fm_{+}(p)\,\bsigma_3\right],
	\label{exp-H0p=}\\[6pt]	
	\bsH_1(x,p)&=\bcU_0(x,p)^{-1}\bcH_1(x,p)\,\bcU_0(x,p-\fK)=
	\chi_\ell(x)\,e^{ix\bcH_0'(p)}\bH_1(x,p)\,e^{-ix\bcH_0'(p-\fK)},
	\label{H=H=H}\\[6pt]
	\bsM_1(p)&=-i\int_0^\ell dx\: \bsH_1(x,p).
	\label{bsM1=}
	\end{align}	
With the help of \eqref{gamma=}, \eqref{bHm-def} -- \eqref{varpi-m}, \eqref{sH-m=}, \eqref{exp-H0p=}, and \eqref{H=H=H}, we then obtain $\gamma_1(x)=-\alpha_1(x)/\alpha_0^2$ and 
	\begin{align}
	\bH_1(x,p)&=-\frac{1}{2}\Big\{\Big[\frac{\alpha_1(x) p(p-\fK)}{\alpha_0^2}+k^2\beta_1(x)\Big]\varpi(p-\fK)^{-1}\bcK+\alpha_1(x)\varpi(p)\bcK^T\Big\},
	\label{bH1=}\\
	\bsH_1(x,p)
	%&=\chi_\ell(x)\,e^{ix\bcH_0'(p)}\bH_1(x,p)\,e^{-ix\bcH_0'(p-\fK)}\\
	&=-\frac{\chi_\ell(x)}{2}\Big\{\Big[\frac{\alpha_1(x) p(p-\fK)}{\alpha_0^2}+k^2\beta_1(x)\Big]
	\varpi(p-\fK)^{-1}\bL_1(x,p)+\alpha_1(x)\varpi(p)\bL_2(x,p)\Big\},
	\label{bsH=4}
	\end{align}
where
	\bea
	\bL_1(x,p)&:=&e^{ix\bcH_0'(p)}\,\bcK\,e^{-ix\bcH_0'(p-\fK)}\nn\\
	&=&\cos[\varphi_-(p)x]\,\bC_{+}(p)+\cos[\varphi_+(p)x]\,\bC_{-}(p)+\nn\\
	&&\sin[\varphi_-(p)x]\,\bS_{+}(p)+\sin[\varphi_+(p)x]\,\bS_{-}(p),	
	\label{L1-def}\\[6pt]
	\bL_2(x,p)&:=&e^{ix\bcH_0'(p)}\,\bcK^T\,e^{-ix\bcH_0'(p-\fK)}\nn\\
	&=&\frac{\tilde\fn(p)\tilde\fn(p-\fK)}{\alpha_0^2}\Big[
	\cos[\varphi_-(p)x]\,\bC_{+}(p)-\cos[\varphi_+(p)x]\,\bC_{-}(p)+\nn\\
	&&\hspace{2.7cm}\sin[\varphi_-(p)x]\,\bS_{+}(p)-\sin[\varphi_+(p)x]\,\bS_{-}(p)\Big],	
	\label{L2-def}\\[6pt]
	\varphi_\pm(p)&:=&\varpi(p)\tilde\fn(p)\pm\varpi(p-\fK)\tilde\fn(p-\fK),
	\label{varphi-pm}
	\\[6pt]
	\bC_{\pm}(p)&:=&\frac{1}{2}\Big[\bcK\pm\frac{\alpha_0^2}{\tilde\fn(p)\tilde\fn(p-\fK)}\,\bcK^T\Big],
	\label{Cpm-def}\\[6pt]
	\bS_{\pm}(p)&=&\check\fm_\pm(p)\,\bI+\check\fm_\mp(p)\,\bsigma_1,
	\label{Spm-def}\\[6pt]
	\check\fm_\pm(p)&:=&\frac{-i\alpha_0}{2}
	\Big[\frac{1}{\tilde\fn(p)}\pm\frac{1}{\tilde\fn(p-\fK)}\Big].
	\label{check-m-pm-def}
	\eea
Substituting \eqref{bsH=4} in \eqref{bsM1=} and using \eqref{L1-def} and \eqref{L2-def}, we find an explicit expression for $\bsH_1(x,p)$. For given choices of $\alpha_1(x)$ and $\beta_1(x)$, we can use this expression to evaluate the integral on the right hand side of \eqref{bsM1=} and read off its matrix elements. This gives $\sM_{1,ab}(p)$. 

{\subsection{{Consistency with reciprocity principle}}}%

{For normal incidence, $\theta_0=0^\circ$ or $\theta_0= {180^\circ}$, Eqs.~\eqref{tau-0p} and \eqref{tau-0m} reproduce the formulas for the reflection and transmission amplitudes for a homogeneous slab which is effectively one dimensional \cite{ptep-2024b,ptep-2026a}. This provides a consistency check on the validity of the formulas we obtained for $\tau_{0\pm}$. In this subsection, we provide a highly nontrivial check on the validity of our formulas for $\tau_{1\pm}$ by showing that they fulfill the requirements of the reciprocity theorem \cite{Rayleigh-1873,potton,Deak-Fulop,Sigwarth}.}

{According to this theorem \cite{Rayleigh-1873}, changing the roles of the source and detector in a scattering process does not affect its outcome. To give a precise quantitative description of this theorem, first we recall that the scattering amplitude $\ff(\theta)$ also depends on the incidence angle $\theta_0$. We make this dependence explicit by adopting the notation $\ff(\theta_0,\theta)$ for the scattering amplitude. We also introduce the symbols $\vec n_0$ and $\vec n$ for the unit vectors in the $x$-$y$ plane that respectively give the directions of the incident wave vector $\vec k_0$ and the position vector $\vec r$ for the detector. It is evident that
	\begin{align}
	&\vec n_0=\cos\theta_0\,\hat\bfe_x+\sin\theta_0\,\hat\bfe_y,
	&& \vec n=\cos\theta\,\hat\bfe_x+\sin\theta\,\hat\bfe_y.
	\label{ns=}
	\end{align}
The reciprocity theorem states that the scattering amplitude is invariant under the transformation $\vec n_0\leftrightarrow -\vec n$, \cite{Landau,ap-2026}. Using \eqref{ns=}, we can express this condition in the form:
	\be
	\ff(\theta_0,\theta)=\left\{\begin{array}{ccc}
	\ff(\theta+ {180^\circ},\theta_0+ {180^\circ})&\for&\cos\theta_0>0~\&~\cos\theta>0,\\
	\ff(\theta- {180^\circ},\theta_0+ {180^\circ})&\for&\cos\theta_0>0~\&~\cos\theta<0,\\
	\ff(\theta+ {180^\circ},\theta_0- {180^\circ})&\for&\cos\theta_0<0~\&~\cos\theta>0,\\
	\ff(\theta- {180^\circ},\theta_0- {180^\circ})&\for&\cos\theta_0<0~\&~\cos\theta<0.
	\end{array}\right.
	\label{reciprocity}
	\ee
In view of \eqref{f=2}, this equation puts strong restrictions on the amplitudes $\tau_{j\pm}$.
}

{For the zeroth-order amplitudes, we can use \eqref{f-zero} -- \eqref{tau-0m} to identify the reciprocity condition \eqref{reciprocity} with the requirement that $\sR^{l/r}(\theta_0)$ and $\sT(\theta_0)$ must be invariant under $\theta_0\to\theta_0\pm {180^\circ}$. To show that this is indeed the case, we observe that, in view of \eqref{m-pm-def}, \eqref{tn-def-0}, and \eqref{T-def} -- \eqref{R-R-def}, $R^{l/r}$ and $T$ are even functions of $p$. This observation together with Eq.~\eqref{Rs-T-zero} and the fact $\sin(\varphi\pm {180^\circ})=-\sin\varphi$ imply 
	\begin{align}
	&\sR^{l/r}(\varphi\pm {180^\circ})=\sR^{l/r}(\varphi),
	&&\sT(\varphi\pm {180^\circ})=\sT(\varphi).
	\label{reciprocity-0}
	\end{align}
This proves that the expressions we found for $\tau_{0\pm}$, namely \eqref{tau-0m} and \eqref{tau-0p}, are indeed invariant under $\theta_0\to\theta_0\pm {180^\circ}$, as required by the reciprocity theorem.} 

{To examine the consistency of the expressions for the amplitudes of the first-order diffracted beams with the reciprocity theorem, first we introduce
	\begin{align}
	&\varsigma(\varphi):=\sin\varphi+\frac{\fK}{k},
	&&\fM_{ab}(\varphi):=\sM_{1,ab}\big(k\,\varsigma(\varphi)\big),
	\label{varsigma=}\\
	&\cR^{l/r}(\varphi):=R^{l/r}\big(k\,\varsigma(\varphi)\big),
	&&\cT(\varphi):=T\big(k\,\varsigma(\varphi)\big),
	\label{cR-cT}
	\end{align}
and let $\ff_1(\theta_0,\theta)$ denote the scattering amplitude corresponding to the first-order diffracted beams. In view of \eqref{f=2}, this is given by
	\begin{align}
	\ff_1(\theta_0,\theta)&:=\tau_{1+}\delta(\theta-\theta_{1+})+\tau_{1-}\delta(\theta-\theta_{1-}).
	\label{f=first}
	\end{align}
It is not difficult to show that
	\begin{align}	
	\ff_1(\theta_0,\theta)&=-\sqrt{2\pi}\,i\times\left\{\begin{array}{ccc}
	t_+^+(\theta_0)\,\delta(\theta-\theta_{1+})&\for&\cos\theta_0>0~\&~\cos\theta>0,\\
	t_-^+(\theta_0)\,\delta(\theta-\theta_{1-})&\for&\cos\theta_0>0~\&~\cos\theta<0,\\
	t_+^-(\theta_0)\,\delta(\theta-\theta_{1+})&\for&\cos\theta_0<0~\&~\cos\theta>0,\\
	t_-^-(\theta_0)\,\delta(\theta-\theta_{1-})&\for&\cos\theta_0<0~\&~\cos\theta<0,
	\end{array}\right.
	\label{f=23} 
	\end{align}
where
	\begin{align}
	&t_+^+(\theta_0):=-\cT(\theta_0)\big[\fM_{11}(\theta_0)+
	\sR^l(\theta_0)\,\fM_{12}(\theta_0)\big],
	\label{tpp=}\\
	&t_+^-(\theta_0):=-\sT(\theta_0)\cT(\theta_0)\,\fM_{12}(\theta_0),\\
	&t^+_-(\theta_0):=  \fM_{21}(\theta_0)+
	\sR^l(\theta_0)\,\fM_{22}(\theta_0) 
	-\cR^l(\theta_0)[\fM_{11}(\theta_0)+
	\sR^l(\theta_{0})\,\fM_{12}(\theta_0)],\\
	&t^-_-(\theta_0):=\sT(\theta_{0})\big[\fM_{22}(\theta_0)
	-\cR^l(\theta_0)\,\fM_{12}(\theta_0)\big],
	\label{tmm=}
	\end{align}
and we have made use of \eqref{tau-1p}, \eqref{tau-1m}, \eqref{varsigma=} -- \eqref{f=first}, and the identities,  
	\begin{align}
	&\cR^{l/r}(\theta_0)=\sR^{l/r}(\theta_{1+}),
	&&\cT(\theta_0)=\sT(\theta_{1+}),
	\label{cRs-cT-def}
	%\\&\sM_{1,ab}\big(k \,\varsigma(\theta)\big)\,\delta(\theta-\theta_{\pm 1})=\sM_{1,ab}(k\,\fs_1)\,\delta(\theta-\theta_{\pm 1}).\nn
	\end{align}
which follow from \eqref{Rs-T-zero} and \eqref{cR-cT}.}

{According to \eqref{sj-def} and \eqref{varsigma=}, 
	\begin{align}
	&\varsigma(\theta_0)=\fs_1=\sin\theta_{1\pm},
	&&|\cos\theta_{1\pm}|=\sqrt{1-\varsigma(\theta_0)^2},
	&&\varsigma(\theta\pm 180^\circ)=-\sin\theta+\tfrac{\fK}{k}.
	\label{eq-36-123}
	\end{align}
These relations together with \eqref{delta-id} imply
	\begin{align}
	\delta(\theta-\theta_{1\pm})&=|\cos\theta_{1+}|\,\delta\big(\sin\theta-\varsigma(\theta_0)\big)~~~\for~~~\pm\cos\theta>0\nn\\
	&=\sqrt{1-\varsigma(\theta_0)^2}\,\delta\big(\sin\theta-\varsigma(\theta_0)\big)~~~\for~~~\pm\cos\theta>0.
	\label{id-659}
	\end{align}
Substituting this equation in \eqref{f=23}, we find
	\begin{align}
	\ff_1(\theta_0,\theta)&=-\sqrt{2\pi}\, i\,\sqrt{1-\varsigma(\theta_0)^2}\,\delta\big(\sin\theta-\varsigma(\theta_0)\big)\times\left\{\begin{array}{ccc}
	t_\pm^+(\theta_0) &\for&\cos\theta_0>0~\&~\pm\cos\theta>0,\\
	t_\pm^-(\theta_0) &\for&\cos\theta_0<0~\&~\pm\cos\theta>0.
	\end{array}\right.
	\label{f=24} 
	\end{align}} %
%	\begin{align}
%	\ff_1(\theta_0,\theta)&=-\sqrt{2\pi}\, i\sqrt{1-\varsigma(\theta_0)^2}\,\delta\big(\sin\theta-\varsigma(\theta_0)\big)\times\left\{\begin{array}{ccc}
%	t_+^+(\theta_0) &\for&\cos\theta_0>0~\&~\cos\theta>0,\\
%	t_-^+(\theta_0) &\for&\cos\theta_0>0~\&~\cos\theta<0,\\
%	t_+^-(\theta_0) &\for&\cos\theta_0<0~\&~\cos\theta>0,\\
%	t_-^-(\theta_0) &\for&\cos\theta_0<0~\&~\cos\theta<0.
%	\end{array}\right.
%	\label{f=24} 
%	\end{align}

{Next, we explore the effects of the transformations 
	\begin{align}
	(\theta_0,\theta)\to(\theta\pm180^\circ,\theta_0+180^\circ)~~~{\rm and}~~~
	(\theta_0,\theta)\to(\theta\pm180^\circ,\theta_0-180^\circ)
	\label{trans-100}
	\end{align}
on $\ff_1(\theta_0,\theta)$ for $\pm\cos\theta>0$. We begin by noting that when the latter condition holds,
	\begin{align}
	&\delta\big(\sin(\theta_0+180^\circ)-\varsigma(\theta\pm180^\circ)\big)=
	\delta\big(\sin(\theta_0-180^\circ)-\varsigma(\theta\pm180^\circ)\big)=
	\delta\big(\sin\theta-\varsigma(\theta_0)\big),\\
	&\sqrt{1-\varsigma(\theta\pm180^\circ)^2}\,\delta\big(\sin(\theta_0+180^\circ)-\varsigma(\theta\pm180^\circ)\big)=|\cos\theta_0|\,\delta\big(\sin\theta-\varsigma(\theta_0)\big),
	\label{id-638a}\\
	&\sqrt{1-\varsigma(\theta\pm180^\circ)^2}\,\delta\big(\sin(\theta_0-180^\circ)-\varsigma(\theta\pm180^\circ)\big)=|\cos\theta_0|\,\delta\big(\sin\theta-\varsigma(\theta_0)\big).
	\label{id-638b}
	\end{align}	
Furthermore, according to \eqref{id-659}, the following identities hold for $\pm\cos\theta>0$.
	\begin{align}
	\delta\big(\sin\theta-\varsigma(\theta_0)\big)t^+_\pm(\theta\pm180^\circ)=
	\delta\big(\sin\theta-\varsigma(\theta_0)\big)t^+_\pm(\theta_{1\pm}\pm180^\circ)~~\for~~\cos\theta_0>0,
	\label{id137a}\\
	\delta\big(\sin\theta-\varsigma(\theta_0)\big)t^-_\pm(\theta\pm180^\circ)=
	\delta\big(\sin\theta-\varsigma(\theta_0)\big)t^-_\pm(\theta_{1\pm}\pm180^\circ)~~\for~~\cos\theta_0<0.
	\label{id137b}
	\end{align}	
For $\pm\cos\theta>0$, \eqref{f=24} and \eqref{id-638a} -- \eqref{id137b} imply
	\begin{align}
	\ff_1(\theta\pm180^\circ,\theta_0+180^\circ)&=-\sqrt{2\pi}\, i\,|\cos\theta_0|\,\delta\big(\sin\theta-\varsigma(\theta_0)\big)\,
	t_-^\mp(\theta_{1\pm}\pm180^\circ)~~\for~~\cos\theta_0>0,
	\nn\\
	\ff_1(\theta\pm180^\circ,\theta_0-180^\circ)&=-\sqrt{2\pi}\, i\,|\cos\theta_0|\,\delta\big(\sin\theta-\varsigma(\theta_0)\big)\,
	t_+^\mp(\theta_{1\pm}\pm180^\circ)~~\for~~\cos\theta_0<0.
	\nn%label{f=25} 
	\end{align}
Comparing these equations with \eqref{f=24} and noting that 
$\sqrt{1-\varsigma(\theta_0)^2}=|\cos\theta_{1+}|$, we infer that $\ff_1$ satisfies the reciprocity condition \eqref{reciprocity} provided that the following (sufficient) conditions hold. 
	 \begin{align}
	|\cos\theta_{1+}|\,t^+_+(\theta_0)= |\cos\theta_0|\,t^-_-(\theta_{1+}+180^\circ)
	~~\for~~\cos\theta_0>0,
	\label{id-rec1}\\
	|\cos\theta_{1+}|\,t^+_-(\theta_0)= |\cos\theta_0|\,t^+_-(\theta_{1-}-180^\circ)
	~~\for~~\cos\theta_0>0,
	\label{id-rec2}\\
	|\cos\theta_{1+}|\,t^-_+(\theta_0)= |\cos\theta_0|\,t^-_+(\theta_{1+}+180^\circ)
	~~\for~~\cos\theta_0<0,
	\label{id-rec3}\\
	|\cos\theta_{1+}|\,t^-_-(\theta_0)= |\cos\theta_0|\,t^+_+(\theta_{1-}-180^\circ)
	~~\for~~\cos\theta_0<0.
	\label{id-rec4}
	\end{align}
	}%

{In the following, we express \eqref{id-rec1} -- \eqref{id-rec4} in terms of $\sM_{1,ab}$. As a first step in this direction, we use \eqref{Rs-T-zero}, \eqref{theta-p-def}, \eqref{sj-def}, \eqref{reciprocity-0}, \eqref{varsigma=}, and \eqref{cRs-cT-def} to establish
	\begin{align}
	\varsigma(\theta_{1\pm}\pm 180^\circ)&=\sin(\theta_{1\pm}\pm 180^\circ)+\tfrac{\fK}{k}=
	-\sin\theta_{1\pm}+\tfrac{\fK}{k}=-\sin\theta_{1+}+\tfrac{\fK}{k}\nn\\
	&=-\sin\theta_0,
	\label{id-varsigma-0}\\	
	\cR^l(\theta_{1\pm}\pm180^\circ)&=R^l\big(k\,\varsigma(\theta_{1\pm}\pm 180^\circ)\big)
	=R^l\big(-k\sin\theta_0\big)=R^l\big(k\sin\theta_0\big)\nn\\
	&=\sR^l(\theta_0),
	\label{id-cRL}\\
	\cT(\theta_{1\pm}\pm180^\circ)&=T\big(k\,\varsigma(\theta_{1\pm}\pm 180^\circ)\big)=
	T(-k\sin\theta_0)=T(k\sin\theta_0)\nn\\
	&=\sT(\theta_0),
	\label{id-cT}\\
	\sR^l(\theta_{1\pm}\pm180^\circ)&=\sR^l(\theta_{1\pm})=\cR^l(\theta_0),
	\label{id-cRL}\\
	\sT(\theta_{1+}+180^\circ)&=\sT(\theta_{1+})=\cT(\theta_0),
	\label{id-sT}\\
	\fM_{ab}(\theta_{1\pm} \pm 180^\circ)&=\sM_{1,ab}\big(k\,\varsigma(\theta_{1\pm} \pm 180^\circ)\big)=\sM_{1,ab} (- k \sin\theta_0 ).
	\label{id-fM}
	\end{align}
In view of \eqref{tpp=} -- \eqref{tmm=}, \eqref{eq-36-123}, and \eqref{id-cRL} -- \eqref{id-sT}, \eqref{id-rec1} -- \eqref{id-rec4} hold provided that $\fM_{ab}$ fulfill the following conditions.
	\begin{align}
	|\cos\theta_{1+}|\;\fM_{11}(\theta_0)&=-|\cos\theta_0| \;\fM_{22}(\theta_{1+}+180^\circ),
	\label{id2-rec1}\\
	|\cos\theta_{1+}|\;\fM_{12}(\theta_0)&=|\cos\theta_0| \;\fM_{12}(\theta_{1\pm}\pm 180^\circ),
	\label{id2-rec2}\\
	|\cos\theta_{1+}|\;\fM_{21}(\theta_0)&=|\cos\theta_0| \;\fM_{21}(\theta_{1-}- 180^\circ),
	\label{id2-rec3}\\
	|\cos\theta_{1+}|\;\fM_{22}(\theta_0)&=-|\cos\theta_0| \;\fM_{11}(\theta_{1-} - 180^\circ).
	\label{id2-rec4}
	\end{align}
This observation reduces the proof of the reciprocity condition \eqref{reciprocity-0} for $\ff_1$ to that of \eqref{id2-rec1} -- \eqref{id2-rec4}. 
}	
	
{Next, we recall that 
	\begin{align}
	&p_0:=k\sin\theta_0, \qquad \varpi(p_0)=k|\cos\theta_0|,\qquad
	k\,\varsigma(\theta_0)=k\sin\theta_0+\fK=p_0+\fK,\\
	&\fM_{ab}(\theta_0)=\sM_{1,ab}\big(k\,\varsigma(\theta_0)\big)=\sM_{1,ab}(p_0+\fK),\\
	&\varpi(p_0+\fK)=\sqrt{k^2-(k\sin\theta_0+\fK)^2}=k\sqrt{1-\sin^2\theta_+}=k|\cos\theta_{1+}|.
	\end{align}
These relations together with \eqref{varsigma=} and \eqref{id-fM} allow us to express  \eqref{id2-rec1} -- \eqref{id2-rec4} in the form
	\begin{align}
	\varpi(p_0+\fK)\;\sM_{1,11}(k\fs_1)&=-\varpi(p_0) \;\sM_{1,22}(-p_0),
	\label{id3-rec1}\\
	\varpi(p_0+\fK)\;\sM_{1,12}(k\fs_1)&=\varpi(p_0) \;\sM_{1,12}(-p_0),
	\label{id3-rec2}\\
	\varpi(p_0+\fK)\;\sM_{1,21}(k\fs_1)&=\varpi(p_0) \;\sM_{1,21}(-p_0),
	\label{id3-rec3}\\
	\varpi(p_0+\fK)\;\sM_{1,22}(k\fs_1)&=-\varpi(p_0) \;\sM_{1,11}(-p_0).
	\label{id3-rec4}
	\end{align}
To show that these equations hold, first we recall that, according to \eqref{varpi-def} and 
\eqref{tn-def-0}, $\varpi$ and $\tilde\fn$ are even functions. This observation together with
\eqref{L1-def} -- \eqref{check-m-pm-def} allow us to verify
	\begin{align}
	&\varphi_\pm(p+\fK)=\pm\varphi_\pm(-p),
	&&\bC_\pm(p+\fK)=\bC_\pm(-p),\nn\\
	&\check\bm_\pm(p+\fK)=\pm\check\bm_\pm(-p),
	&&[\bL_a(x,p+\fK)]_{11}= -[\bL_a(x,-p)]_{22},\nn\\
	&[\bL_a(x,p+\fK)]_{12}= [\bL_a(x,-p)]_{12},
	&&[\bL_a(x,p+\fK)]_{21}= [\bL_a(x,-p)]_{21},\nn
	\end{align}
for  both $a=1$ and $a=2$. We can use these relations and \eqref{bsH=4} to prove
	\begin{align}
	&\varpi(p)\sH_{11}(p+\fK)=-\varpi(p+\fK)\sH_{22}(-p),\nn\\
	&\varpi(p)\sH_{12}(p+\fK)=\varpi(p+\fK)\sH_{12}(-p),\nn\\
	&\varpi(p)\sH_{21}(p+\fK)=\varpi(p+\fK)\sH_{21}(-p).\nn
    \end{align}	
Equations \eqref{id3-rec1} -- \eqref{id3-rec2} follow from these identities and \eqref{bsM1=}. Since \eqref{id3-rec1} -- \eqref{id3-rec2} are equivalent to \eqref{id2-rec1} -- \eqref{id2-rec4}, this completes the proof that $\ff_1$ satisfies the reciprocity condition~\eqref{reciprocity-0}.
}\\
 
\section{Application to {zeroth- and} first-order diffracted beams}
\label{S4}

{Having obtained explicit formulas for the zeroth- and first-order diffracted beam amplitudes, we are now in a position to apply these formulas to specific gratings of the form~\eqref{model-1}. Because of their wider range of applications,} we confine our attention to nonmagnetic diffraction gratings of the form~\eqref{model-1}. This corresponds to situations where $\fb_n=0$ for all $n\geq 0$. We consider the scattering of TE and TM waves by such a grating separately.

  For TE waves, setting $\fb_n=0$ in \eqref{alpha-beta-n=}, we have $\alpha_n=\delta_{0n}$ and $\beta_n=\delta_{0n}+\fa_n$. These relations together with \eqref{tn-def-0}, \eqref{bsM1=}, \eqref{bsH=4}, and \eqref{check-m-pm-def} imply
	\bea
	&&\bsH_1(x,p)=-\frac{1}{2}\,\chi_\ell(x) k^2\,\varpi(p-\fK)^{-1}\fa_1(x)\,\bL_1(x,p) ,
	\label{bsH=nonmag}\\
	&&\bsM_1(p)=\frac{i k^2}{2}\,\varpi(p-\fK)^{-1}\int_0^\ell dx\: 
	\fa_1(x)\bL_1(x,p),
	\label{sM1=nonmag-TE}\\
	&&\tilde\fn(p):=\pm\,\sqrt{1+\frac{\fa_0}{1-p^2/k^2}},
	\label{tn=nonmag}
	\eea
Substituting \eqref{tn=nonmag} in \eqref{varphi-pm} -- \eqref{check-m-pm-def} and using the resulting equation together with \eqref{varpi-def}, \eqref{L1-def}, and \eqref{sM1=nonmag-TE}, we can determine the explicit form of $\bsM_1(p)$ and read off $\sM_{1,ab}(p)$. 

Similarly, for TM waves, substituting $\fb_n=0$ in \eqref{alpha-beta-n=}, we find
	\begin{align}
	&\alpha_n=\delta_{0n}+\fa_n, 
	&&\beta_n=\delta_{0n}.
	\label{nonmagnetic}
	\end{align} 
In this case, we can use \eqref{tn-def-0} to show that \eqref{tn=nonmag} still holds. Furthermore, substituting \eqref{nonmagnetic} in \eqref{bsM1=} and \eqref{bsH=4}, we obtain
	\bea
	&&\bsH_1(x,p)=-\frac{1}{2}\,\chi_\ell(x)\left[
	\frac{p(p-\fK)\,\fa_1(x)\,\bL_1(x,p)}{(\fa_0+1)^2\varpi(p-\fK)}
	+\varpi(p)\fa_1(x)\bL_2(x,p)\right],
	\label{bsH=nonmag-TM}\\
	&&\bsM_1(p)=\frac{i}{2}\left[
	\frac{p(p-\fK)}{(\fa_0+1)^2\varpi(p-\fK)}\int_0^\ell dx\: \fa_1(x)\bL_1(x,p)+
	\varpi(p)\int_0^\ell dx\:\fa_1(x)\bL_2(x,p)\right].
	\label{sM1=nonmag-TM}
	\eea
Again we can use \eqref{varpi-def}, \eqref{L1-def} -- \eqref{check-m-pm-def}, \eqref{tn=nonmag}, and \eqref{sM1=nonmag-TM} to calculate $\sM_{1,ab}(p)$ for TM waves. 

As a concrete example, consider the scattering of right-incident TE and TM waves that are scattered by a nonmagnetic grating given by \eqref{model-1}, $N=1$, $\fb_0=\fb_1=0$, and
	\be
	\fa_1(x):=\zeta\, e^{-\kappa (\ell-x)},
	\label{eg2}
	\ee
where $\zeta$ and $\kappa$ are positive real parameters. This corresponds to a relative permittivity profile of the form
	\be
	\hat\varepsilon(x,y)=1+\chi_\ell(x)\left[\fa_0+\zeta\, e^{-\kappa (\ell-x)} e^{i\fK y}\right],
	\label{eg2-epsilon}
	\ee
which generalizes Berry's grating \eqref{berry1}. Demanding that $\zeta$ and $\fa_0$ satisfy
	\be
	\zeta<\IM(\fa_0)\ll 1<\RE(\fa_0)+1,
	\label{lossy}
	\ee
we can ensure that {condition \eqref{condi} holds, the real part of $\hat\varepsilon$ takes values that are not smaller than 1, and its imaginary part takes nonnegative values. In particular, this grating is made of nonexotic lossy material.}
		
Substituting \eqref{eg2} in \eqref{sM1=nonmag-TE} and \eqref{sM1=nonmag-TM}, we can perform the integrals on the right-hand side of these equations analytically, and find explicit closed-form expressions for $\sM_{1,ab}$ and consequently $\tau_{1\pm}$. Because the resulting expressions are lengthy, we provide their graphical representations for certain realistic values of the system's physical parameters.

\begin{figure}
        \begin{center}
        \includegraphics[scale=.25]{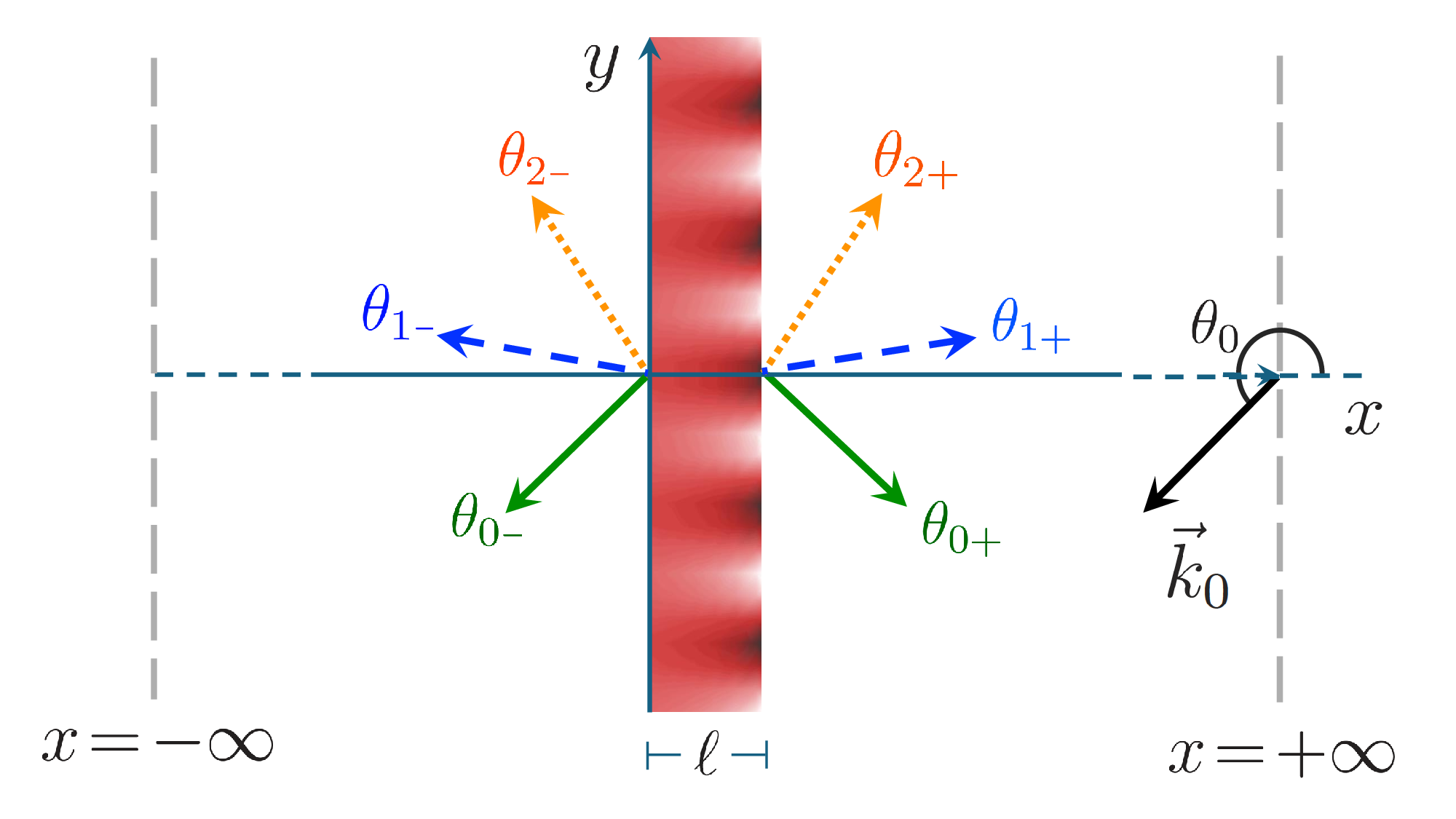} ~~~~~
        \includegraphics[scale=.5]{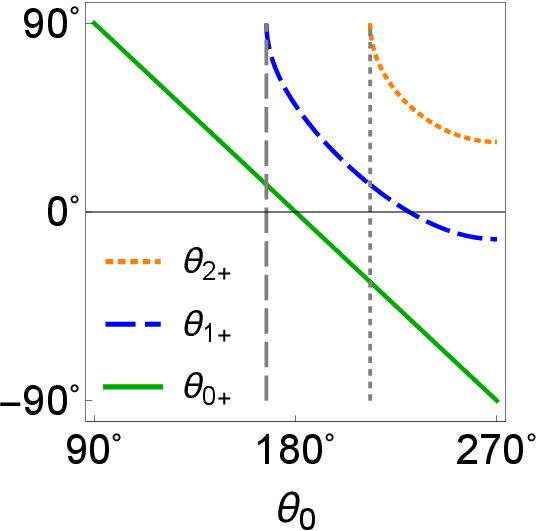} 
        \caption{The left panel shows a schematic view of the scattering of a right-incident TE or TM wave by a nonmagnetic diffraction grating given by \eqref{eg2-epsilon}, \eqref{specs1}, and \eqref{K=eg}. $\theta_0$ and $k$ are chosen such that $J=2$. Arrows labelled by the angles $\theta_{j\pm}$ represent the diffracted beams. A density plot of $|\hat\epsilon(x,y)|$ is used to color the grating in shades of red. The right panel gives the plots of $\theta_{+j}$ for $j\in\{0,1,2\}$ as functions of $\theta_0$ for waves with wavelength $1.55\,\mu{\rm m}$. The first- and second-order diffracted beams are generated provided that $\theta_0$ respectively exceeds $167.0^\circ$ and $213.4^\circ$. The vertical dashed and dotted gray lines mark these values of $\theta_0$.}
        \label{fig2}
        \end{center}
        \end{figure}

Figure~\ref{fig2} provides a schematic view of the scattering of a right-incident TE or TM wave by a nonmagnetic grating made of an InGaAsP slab whose relative permittivity is tailored  to have of a profile of the form \eqref{eg2-epsilon} with
	\begin{align}
	&\ell=2\:\mu{\rm m},
	&\fa_0=9.57+i 0.05,
	&&\zeta=0.03,
	&&\kappa=1\:\mu{\rm m}^{-1},
	%&&\fK=\pi~\mu{\rm m}^{-1}=3.14~\mu{\rm m}^{-1}.
	\label{specs1}
	\end{align}
and 	
	\be
	\fK=\frac{2\pi}{L}=3.14\,\mu{\rm m}^{-1},
	\label{K=eg}
	\ee
which corresponds to taking  the period $L:=2\pi/\fK$ of the variations of the permittivity profile of the grating along the $y$ direction to be $L=2\,\mu{\rm m}$. The wavelength of the incident wave is taken as $\lambda=1.55\,\mu{\rm m}$ which corresponds to the incident wavenumber:
	\be
	k=\frac{2\pi}{\lambda}=4.05\, \mu{\rm m}^{-1}.
	\label{k=eg}
	\ee
 For these values of $k$ and $\fK$, \eqref{theta-p-def} and \eqref{theta-J} give 
 	\begin{align}
	&J\in\{0,1, 2\},
	&&
	\theta_J=\left\{\begin{array}{ccc}
	~90.0^\circ&\for&J=0,\\
	~13.0^\circ&\for&J=1,\\
	-33.4^\circ&\for&J=2.
	\end{array}\right.
	\label{theta-J-eg}
	\end{align}
 In view of \eqref{condi-94}, \eqref{theta-J-eg}, and the fact that for a right-incident wave $\theta_0>90^\circ$, the first- and second-order diffracted beams arise when $\theta_0$ exceeds $167.0^\circ$ and $213.4^\circ$, respectively. This is clearly shown in the plots of $\theta_{j+}$ as functions of $\theta_0$, which are also given in Fig.~\ref{fig2}. 
  	
\begin{figure}
        \begin{center}
        \includegraphics[scale=.8]{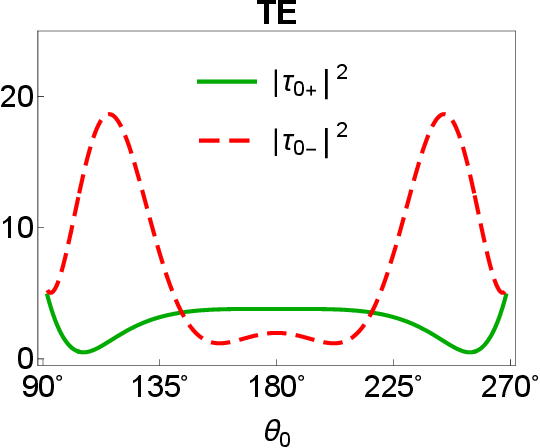} ~~~~~~~~~
        \includegraphics[scale=.8]{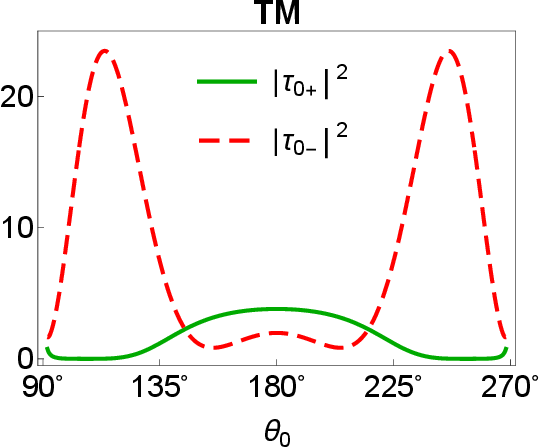} \\[12pt]
        \includegraphics[scale=.6]{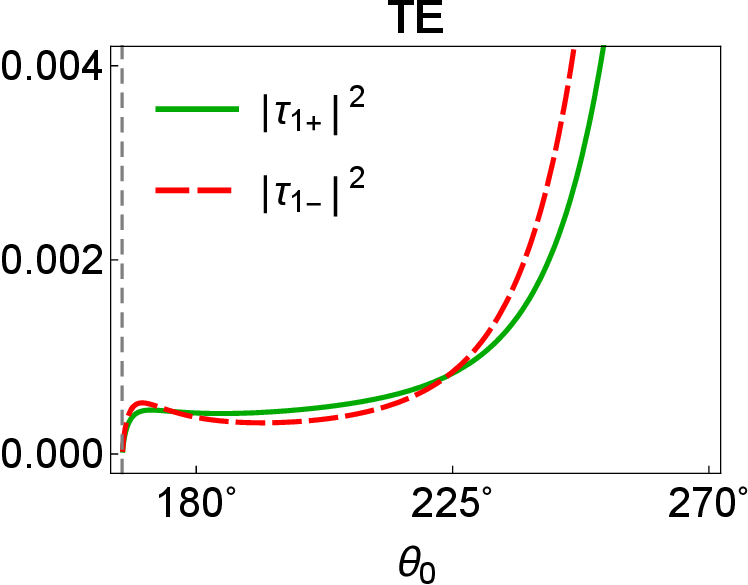} ~~~~~~~~~
        \includegraphics[scale=.6]{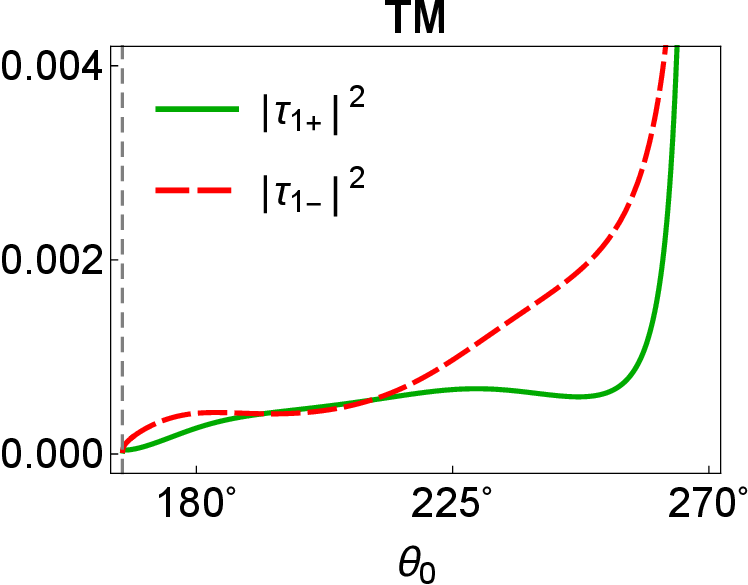} \\[12pt]
	\includegraphics[scale=.6]{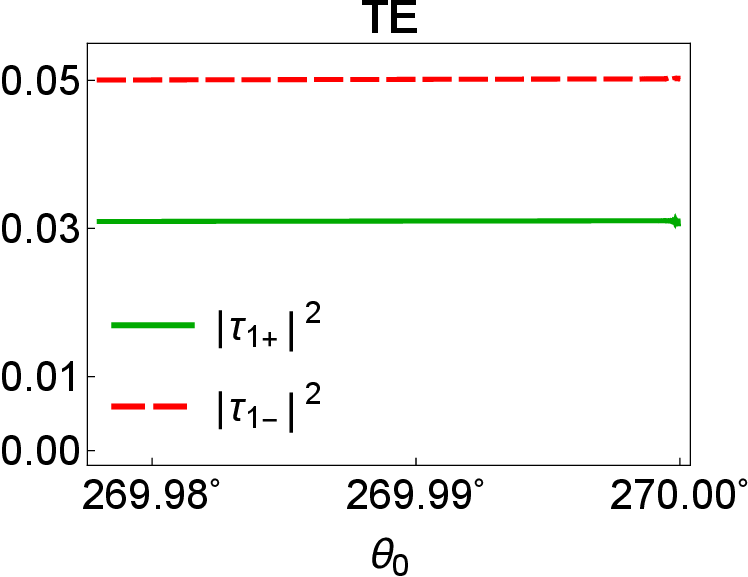} ~~~~~~~~~
	\includegraphics[scale=.6]{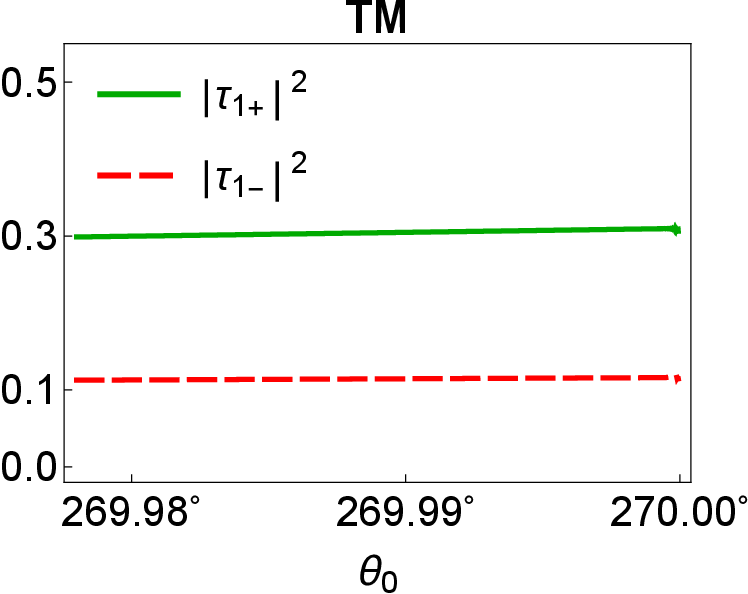}
        \caption{Plots of $|\tau_{0\pm}|^2$ and $|\tau_{1\pm}|^2$ as functions of $\theta_0$ for the scattering of right-incident TE and TM waves with wavelength $1.55~\mu{\rm m}$ by a nonmagnetic diffraction grating given by \eqref{eg2-epsilon} {with $\ell=2\:\mu{\rm m}$, $\fa_0=9.57+i 0.05$, $\zeta=0.03$, $\kappa=1\:\mu{\rm m}^{-1}$, and $\fK=3.14~\mu{\rm m}^{-1}$. Because $\cos\theta_0<0$, $\tau_{j+}$ and $\tau_{j-}$ respectively correspond to the reflected and transmitted beams.} The dashed vertical line marks $\theta_0=167.0^\circ$, i.e., the minimum value of $\theta_0$ for which the first-order diffracted beams are present. At grazing angle, where $\theta_0$ approaches $270^\circ$, $|\tau_{1\pm}|^2$ tend to finite values which for TM waves are about an order of magnitude larger than those for TE waves. These are the largest values attained by $|\tau_{1\pm}|^2$ for both TE and TM waves. }
        \label{fig3}
        \end{center}
        \end{figure}

Figure~\ref{fig3} shows the plots of $|\tau_{0\pm}|^2$ and $|\tau_{1\pm}|^2$ as functions of  $\theta_0$ for right-incident TE and TM waves with wavelength $\lambda=1.55\,\mu{\rm m}$ scattered by a nonmagnetic grating given by \eqref{eg2-epsilon}, \eqref{specs1}, and \eqref{K=eg}. {To generate these plots, first we used \eqref{m-pm-def}, \eqref{tn-def-0}, and \eqref{T-def} -- \eqref{Rs-T-zero} to determine $\sT(\varphi)$ and $\sR^{l/r}(\varphi)$. Next, we employed \eqref{L1-def} -- \eqref{check-m-pm-def}, \eqref{sM1=nonmag-TE}, and \eqref{sM1=nonmag-TM} to calculate the matrix $\bsM_1(p)$ for TE and TM waves. Identifying the entries of $\bsM_1(p)$ for each of these waves and substituting the resulting expressions together with those of $\sT(\varphi)$ and $\sR^{l/r}(\varphi)$ in \eqref{tau-0p}, \eqref{tau-0m}, \eqref{tau-1p}, and \eqref{tau-1m}, we obtain and plot $|\tau_{0\pm}|^2$ and $|\tau_{1\pm}|^2$.}

 {As shown in Fig.~\ref{fig3}, when} $\theta_0$ tends to $270^\circ$ (grazing angle), $|\tau_{1\pm}|^2$ take much larger values. A careful analysis of the behavior of $|\tau_{1\pm}|^2$ for $\theta_0$ shows that they tend to finite values as $\theta_0$ approaches $270^\circ$. The graphs of $|\tau_{1\pm}|^2$ for $\theta_0$ between $269.98^\circ$ and $270.00^\circ$, also shown in Fig.~\ref{fig3}, provide a graphical confirmation of this effect. We have also examined the plots of $|\tau_{0\pm}|^2$ for other values of $\fK$ and found that the qualitative behavior depicted in Fig.~\ref{fig3} persists. In particular, at grazing angle, $|\tau_{1+}|^2<|\tau_{1-}|^2$ for TE waves. For TM waves incident at grazing angle, depending on the value of $\fK$, $|\tau_{1+}|^2$ can take larger or smaller values than $|\tau_{1-}|^2$.  
        
\begin{figure}
        \begin{center}
        \includegraphics[scale=.6]{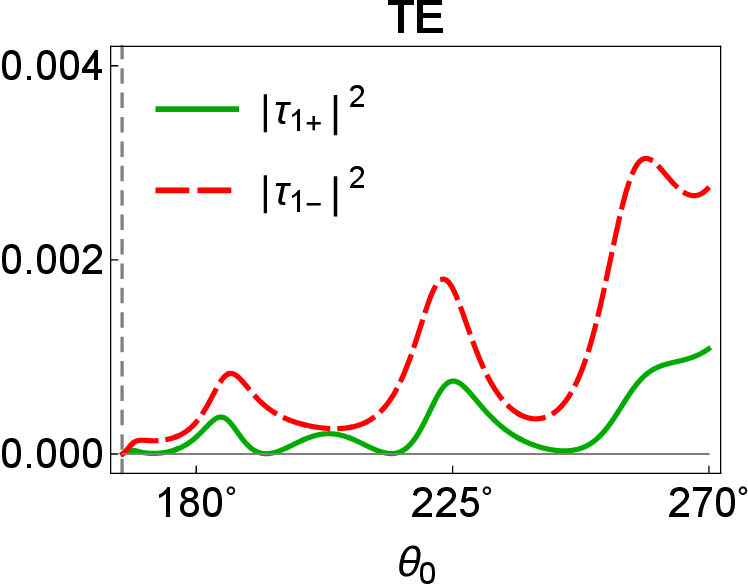} ~~~~~~~
        \includegraphics[scale=.6]{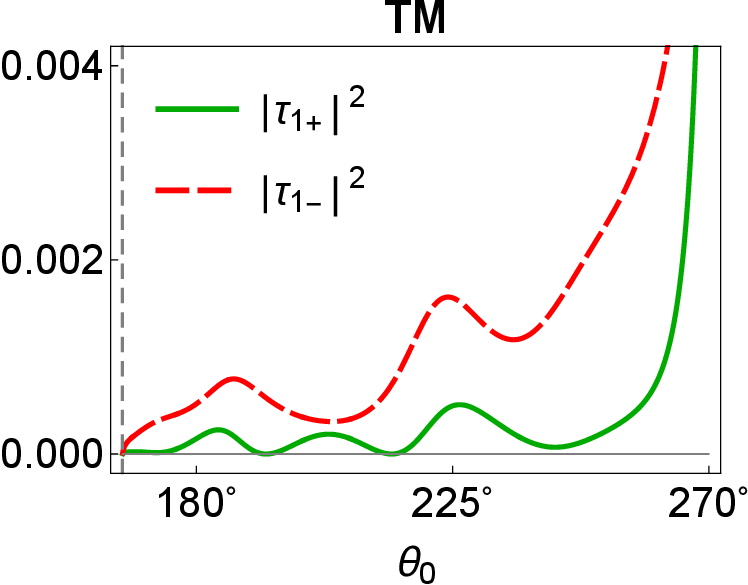} 
        \caption{Plots $|\tau_{1\pm}|^2$ as functions of $\theta_0$ for the scattering of right-incident TE and TM waves with wavelength $1.55~\mu{\rm m}$ by a nonmagnetic diffraction grating {given by \eqref{eg2-epsilon} with $\ell=10\:\mu{\rm m}$, $\fa_0=9.57+i 0.05$, $\zeta=0.03$, $\kappa=1\:\mu{\rm m}^{-1}$, and $\fK=3.14~\mu{\rm m}^{-1}$.} The dashed vertical line marks $\theta_0=167.0^\circ$, i.e., the minimum value of $\theta_0$ for which the first-order diffracted beams are present.}
        \label{fig4}
        \end{center}
        \end{figure}

Examining the plots of $|\tau_{1\pm}|^2$ for different values of $\ell$, we find that decreasing the grating thickness does not produce any qualitative change in the behavior of $|\tau_{1\pm}|^2$, whereas increasing it has a noticeable effect. In particular, for sufficiently thick gratings, $\tau_{1+}$ develops a finite set of isolated zeros for both TE and TM waves. This means that there is a finite set of values of the incidence angle at which the reflected first-order beam is absent while the transmitted first-order beam is present.  Figure~\ref{fig4} shows the graphs of $|\tau_{1\pm}|^2$ for a nonmagnetic grating with the same physical parameters as the one considered in Fig.~\ref{fig3} except that it is $10~\mu{\rm m}$ thick, i.e., $\ell=10~\mu{\rm m}$.
       
%We close this section by recalling the fact that our analytic calculation of $\tau_{1\pm}$ is valid also for gratings \eqref{model-1} with $N>1$. 

{Our analytic expressions for $\tau_{0\pm}$ and $\tau_{1\pm}$ allows us to carry out a similar analysis for magnetic gratings with permittivity and permeabilities of the form \eqref{model-1}. To realize these, however, one would need a carefully engineered metamaterial slab. We do not elaborate on the outcomes of such an analysis, as they will depend on the values of the physical parameters of the metamaterial slab in question.}

\section{Summary and concluding remarks} 
\label{S5}

When a plane wave is incident upon a planar slab with periodic inhomogeneities along one of the tangential directions to its boundaries, the scattered wave consists of a finite set of reflected and transmitted beams. This phenomenon is known as diffraction, and such a slab is called a diffraction grating. In this article, we consider a large class of diffraction gratings whose scattering problem can be solved exactly. In other words, we outline a method for computing the complex amplitudes of the diffracted beams that involves only a finite number of algebraic operations and the evaluation of finitely many definite integrals.

The basic tool we employ to establish the exact solvability of this class of diffraction gratings is an approach to stationary scattering that maps scattering problems to those of determining the dynamics of an associated effective quantum system. The central ingredient of this approach
%, which we call the “dynamical formulation of stationary scattering,” 
is the concept of the fundamental transfer matrix. This plays a role analogous to that of the scattering matrix in the textbook treatment of stationary scattering, as it encodes the scattering properties of the system. The connection with an effective quantum dynamics yields a Dyson series expansion for the fundamental transfer matrix. The determination of the latter, however, is not sufficient for evaluating the scattering amplitude. This requires solving a pair of linear (integral) equations that, in general, admit only formal series solutions.

For the class of diffraction gratings considered in this article, both the Dyson series for the fundamental transfer matrix and the series solutions of the equations determining the scattering amplitude truncate. This allows for an exact calculation of the complex amplitudes of the diffracted beams.

Our results apply to both TE and TM waves with arbitrary angles of incidence; that is, they cover both the Bragg and Raman–Nath regimes \cite{Moharam-1978}. Their application is naturally simpler in the Bragg regime, where the number of diffracted beams is small. Nevertheless, the method can also be applied consistently to compute the amplitudes of higher-order beams in the Raman–Nath regime. Although the resulting analytic calculations become extremely involved for second- and higher-order beams, {they can be performed using computer-aided symbolic calculations. They may also} provide a basis for developing more efficient numerical methods for computing the corresponding amplitudes.

\subsection*{Acknowledgements} 
This work has been supported by the Scientific and Technological Research Council of T\"urkiye (T\"UB\.{I}TAK) in the framework of the project 123F180 and by Turkish Academy of Sciences (T\"UBA). {A.~M.\ wishes to express his gratitude to Alireza Qaiumzadeh, Alphan Sennaro\u{g}lu, and Stefan Rotter for helpful discussions.}

\section*{Appendix A: Properties of $\widehat S$}

Consider the linear operators $\widehat\Pi_k$ and $\widehat S$ defined by \eqref{Pi-k-def} and \eqref{S-def}, respectively. We can view these operators as linear operators acting in either of $\sF$ or $\sF^2$; they satisfy \eqref{Pi-k} and \eqref{shift} for both $\phi\in\sF$ and $\phi\in\sF^2$. 

Given a positive integer $m$, we can use \eqref{Pi-k-def} and \eqref{shift} to conduct the following calculation.
	\begin{align}
	\widehat\Pi_k\widehat S^m\widehat\Pi_k&=
	\int_{-k}^k dp\int_{-k}^k dp'\: |p'\kt\br p'|\widehat S^m|p\kt\br p|\nn\\
	&=\int_{-k}^k dp\int_{-k}^k dp'\: |p'\kt\br p'-m\fK|p\kt\br p|\nn\\
	&=\int_{-k}^k dp\int_{-k}^k dp'\: \delta(p'-m\fK-p)|p'\kt\br p|\nn\\
	&=\int_{-k}^k dp\: \Theta(k-|p+m\fK|) |p+m\fK\kt\br p|\nn\\
	&=\Theta(\tfrac{2k}{\fK}-m)\int_{-k}^{k-m\fK}dp\:|p+m\fK\kt\br p|\nn\\
	&=\Theta(\tfrac{2k}{\fK}-m)\int_{-k+m\fK}^{k}dq\:|q\kt\br q-m\fK|\nn\\
	&=\widehat\Lambda_m\,\widehat S^m,
	\label{app-e1}
	\end{align}
where $\Theta$ is the Heaviside step function, i.e., $\Theta(x)=1$ for $x\geq 0$, and $\Theta(x)=0$ for $x< 0$, and 
	\be
	\widehat\Lambda_m:=\Theta(\tfrac{2k}{\fK}-m)\int_{-k+m\fK}^{k}dq\:|q\kt\br q|.
	\label{Lambda-m-def}
	\ee
Equation~\eqref{terminate} is a direct consequence of \eqref{app-e1} and \eqref{Lambda-m-def}. 

In view of \eqref{Pi-k-def} and \eqref{Lambda-m-def}, $\widehat\Lambda_m$ satisfies  			\begin{align}
&\widehat\Pi_k\widehat\Lambda_m=\widehat\Lambda_m\widehat\Pi_k=\widehat\Lambda_m.
	\label{id-202}
	\end{align} 
Furthermore, because $\widehat\Pi_k^2=\widehat\Pi_k$, \eqref{app-e1} and \eqref{id-202} imply
	\be
	\widehat\Pi_k\widehat S^m\widehat\Pi_k=
	\widehat\Lambda_m \widehat S^m\widehat\Pi_k=
	\widehat\Pi_k\widehat\Lambda_m\,\widehat S^m\widehat\Pi_k=
	\widehat\Lambda_m\,\widehat\Pi_k\widehat S^m\widehat\Pi_k.
	\label{app-e2}
	\ee
	
Next, let $u:\R\to\C$ be a function, and $n$ be a positive integer. Then, by virtue of \eqref{shift}, \eqref{app-e1}, \eqref{Lambda-m-def}, and \eqref{app-e2}, we have  $\widehat\Lambda_{m}\widehat\Lambda_{m+n}=\widehat\Lambda_{m+n}$ and
	\begin{align}
	\widehat\Pi_k\widehat S^m u(\widehat p)\,\widehat\Pi_k \widehat S^n \widehat\Pi_k
	&=\widehat\Pi_k u(\widehat p-m\fK\widehat I\,)\,
	\widehat S^m\widehat\Pi_k\widehat S^{n}\widehat\Pi_k\nn\\
	&=u(\widehat p-m\fK\widehat I\,)(\widehat\Pi_k \widehat S^m \widehat\Pi_k)
	\widehat S^{n}\widehat\Pi_k\nn\\
	&=u(\widehat p-m\fK\widehat I\,)\,\widehat\Lambda_m\widehat S^{m+n}\widehat\Pi_k
	\nn\\
	&=u(\widehat p-m\fK\widehat I\,)\,\widehat\Lambda_{m}\,
	\widehat\Pi_k\widehat S^{m+n}\widehat\Pi_k\nn\\
	&=u(\widehat p-m\fK\widehat I\,)\,\widehat\Lambda_{m}
	\widehat\Lambda_{m+n}\,\widehat S^{m+n}\nn\\
	&=u(\widehat p-m\fK\widehat I\,)\,\widehat\Lambda_{m+n}\,\widehat S^{m+n}\nn\\
	&=u(\widehat p-m\fK\widehat I\,)\,\widehat\Pi_k\,\widehat S^{m+n}\widehat\Pi_k.
	\nn
	\end{align}
This proves \eqref{id-101}.

\section*{Appendix B: Derivation of \eqref{A-left-2} -- \eqref{B-right-2}}

To obtain an explicit expression for $B_-^r$, we substitute \eqref{M22=2-inv} in the second equation in \eqref{B-right}. Making use of \eqref{tdelta} and \eqref{id-1}, we can then show that
	\begin{align}
	&B^r_-={T}(p_0)\sum_{j=0}^{\lfloor\!\lfloor 2k/\fK \rfloor\!\rfloor}  \big[\widehat\sN_{22}-
	R^l(\widehat p)\widehat\sN_{12}\big]^j 
	\check\delta_{p_0},
	\label{B-right=}\\
	&\begin{aligned}
	B^r_-(p)&= 2\pi\varpi_0 {T}(p_0)\sum_{j=0}^{\lfloor\!\lfloor 2k/\fK \rfloor\!\rfloor}
	\br p|\big[\widehat\sN_{22}-R^l(\widehat p)\widehat\sN_{12}\big]^j|p_0\kt\\
	&=2\pi\varpi_0 {T}(p_0)\Big\{\delta(p-p_0)+
	\sum_{j=1}^{\lfloor\!\lfloor 2k/\fK \rfloor\!\rfloor}
	\br p|\big[\widehat\sN_{22}-R^l(\widehat p)\widehat\sN_{12}\big]^j|p_0\kt\Big\}
	\end{aligned}
	\label{B-right=p}
	\end{align} 
Setting $p=p_1:=k\sin\theta$ in the latter equation and employing \eqref{tdelta}, we arrive at \eqref{B-right-2}. 

Next, we use \eqref{M=MM}, \eqref{sN-ab}, and \eqref{T-def} -- \eqref{R-R-def}, to show that
	\begin{align}
	\widehat M_{11}&=
	M_{0,11}(\widehat p)(\widehat I-\widehat\sN_{11})-
	T(\widehat p)^{-1}R^r(\widehat p)\widehat\sN_{21}\nn\\
	&=T(\widehat p)^{-1}\left[S(\widehat p)(\widehat I-\widehat\sN_{11})-R^r(\widehat p)
	\widehat\sN_{21}\right],
	\label{M11=B}\\
	\widehat M_{12}&=
	T(\widehat p)^{-1}R^r(\widehat p)(\widehat I-\widehat\sN_{22})
	-M_{0,11}(\widehat p)\widehat\sN_{12}\nn\\
	&=T(\widehat p)^{-1}\left[R^r(\widehat p)(\widehat I-\widehat\sN_{22})
	-S(\widehat p)\widehat\sN_{12}\right],
	\label{M12=B}\\
	\widehat M_{21}&=
	-T(\widehat p)^{-1}\big[R^l(\widehat p)(\widehat I-\widehat\sN_{11})
	+\widehat\sN_{21}\big],
	\label{M21=B}
	\end{align}
where
	\be
	S(p):=\frac{M_{0,11}(p)}{M_{0,22}(p)}=T(p)M_{0,11}(p)=T(p)^2-R^l(p)R^r(p).
	\label{S-def-id}
	\ee
The last equality in \eqref{S-def-id} follows from the fact that $\det\bM_0(p)=1$, \cite{jpa-2025}.  
	
Substituting \eqref{B-right=} and \eqref{M12=B} in the second equation in \eqref{As=} and using \eqref{condi-sN} and \eqref{S-def-id}, we find after surprising cancellations:
	\begin{align}	
	&\begin{aligned}
	A_+^r=&\:R^r(p_0)\check\delta_{p_0}
	-T(p_0)T(\widehat p)\widehat\sN_{12}\!\!\sum_{j=1}^{\lfloor\!\lfloor 2k/\fK \rfloor\!\rfloor}\!\!
	\big[\widehat\sN_{22}-R^l(\widehat  p)\widehat\sN_{12}\big]^{j-1}
	\check\delta_{p_0},
	\end{aligned}
	\label{A-right}\\
	&\begin{aligned}
	A_+^r(p)=&\:2\pi  R^r(p_0)\delta(p-p_0)-
	2\pi\varpi_0\,T(p_0)T(p)\!\!\sum_{j=1}^{\lfloor\!\lfloor 2k/\fK \rfloor\!\rfloor}\! 
	\br p| \widehat\sN_{12}\big[\widehat\sN_{22}-R^l(\widehat  p)\widehat\sN_{12}\big]^{j-1}|p_0\kt.
	\end{aligned}
	\label{A-right=p}
	\end{align}
Similarly, using \eqref{tdelta}, \eqref{As=}, \eqref{Bs=}, \eqref{sN-ab}, \eqref{M22=2} -- \eqref{R-R-def}, \eqref{M11=B} -- \eqref{M21=B}, and \eqref{S-def-id}, we obtain
	\begin{align}
	&\begin{aligned}
	B^l_-	=&\:R^l(p_0)\check\delta_{p_0}+\sum_{j=1}^{\lfloor\!\lfloor 2k/\fK \rfloor\!\rfloor}
	\Big\{\big[\widehat\sN_{22}-R^l(\widehat p)\widehat\sN_{12}\big]^jR^l(\widehat p)\,+\\
	&\hspace{3cm}\big[\widehat\sN_{22}-R^l(\widehat p)\widehat\sN_{12}\big]^{j-1}
	\big[\widehat\sN_{21}-R^l(\widehat p)\widehat\sN_{11}\big]\Big\}\check\delta_{p_0},
	\end{aligned}
	\label{B-left=}\\
	&\begin{aligned}
	B^l_-(p)=&\:2\pi\varpi_0\, R^l(p_0)\delta(p-p_0)+
	2\pi\varpi_0\!\!\sum_{j=1}^{\lfloor\!\lfloor 2k/\fK \rfloor\!\rfloor}\!\!
	\Big\{R^l(p_0)\br p|\big[\widehat\sN_{22}-R^l(\widehat p)\widehat\sN_{12}\big]^j|p_0\kt+\\
	&\br p|\big[\widehat\sN_{22}-R^l(\widehat p)\widehat\sN_{12}\big]^{j-1}
	\big[\widehat\sN_{21}-R^l(\widehat p)\widehat\sN_{11}\big]|p_0\kt\Big\},
	\end{aligned}
	\label{B-left=p}
	\end{align}
	\begin{align}
	&\begin{aligned}
	A^l_+&=T(p_0)\check\delta_{p_0}-T(\widehat p)
	\big[\widehat\sN_{11}+R^l(p_0)\widehat\sN_{12}\big]-T(\widehat p)
	\!\!\sum_{j=2}^{\lfloor\!\lfloor 2k/\fK \rfloor\!\rfloor}\!\! \Big\{R^l(p_0)
	\widehat \sN_{12}\big[\widehat\sN_{22}-R^l(\widehat p)\widehat\sN_{12}\big]^{j-1}
	\\
	&~~~+\widehat\sN_{12}
	\big[\widehat\sN_{22}-R^l(\widehat p)\widehat\sN_{12}\big]^{j-2}
	\big[\widehat\sN_{21}-R^l(\widehat p)\widehat\sN_{11}\big]
	\Big\}\check\delta_{p_0},
	\end{aligned}
	\label{A-left=}\\
	&\begin{aligned}
	A^l_+(p)&=2\pi\varpi_0T(p_0)\delta(p-p_0)-2\pi\varpi_0 T(p)
	\big[\br p|\widehat\sN_{11}|p_0\kt+R^l(p_0)\br p|\widehat\sN_{12}|p_0\kt\big]\\
	&~~-2\pi\varpi_0 T(p)
	\!\!\sum_{j=2}^{\lfloor\!\lfloor 2k/\fK \rfloor\!\rfloor}\!\! \Big\{R^l(p_0)
	\br p|\widehat \sN_{12}\big[\widehat\sN_{22}-R^l(\widehat p)\widehat\sN_{12}\big]^{j-1}
	|p_0\kt\\
	&
	\hspace{3.3cm}+\br p|\widehat\sN_{12}
	\big[\widehat\sN_{22}-R^l(\widehat p)\widehat\sN_{12}\big]^{j-2}
	\big[\widehat\sN_{21}-R^l(\widehat p)\widehat\sN_{11}\big]|p_0\kt\Big\}.
	\end{aligned}
	\label{A-left=p}
	\end{align}
	
Next, we note that, according to Eq.~\eqref{f=}, to evaluate the scattering amplitude we only need the expressions for $A^l_+(k\sin\theta)$ and $B^r_-(k\sin\theta)$ for the values of $\theta$ such that $\cos\theta$ and $\cos\theta_0$ have the same sign. If we set $p=p_1=k\sin\theta$ in \eqref{B-right=p}, \eqref{A-right=p},  \eqref{B-left=p}, and \eqref{A-left=p} and use \eqref{tdelta} and the identity,
	\be
	\delta(p_1-p_0)=\delta[k(\sin\theta-\sin\theta_0)]=
	\frac{\delta(\theta-\theta_0)}{k|\cos\theta_0|}~~~\for~~~
	\cos\theta\,\cos\theta_0>0,
	\ee
we arrive at \eqref{B-right-2}, \eqref{A-right-2},  \eqref{B-left-2}, and \eqref{A-left-2}, respectively.
%\pagebreak
%
{\section*{Appendix C: Diffracted beam intensities}}

{The intensity of an electromagnetic wave is, by definition, the magnitude of its time-averaged Poynting vector. For the time-harmonic TE and TM waves reaching the detectors, the latter has the form 
	\be
	\br\vec S\kt:=\frac{c_b}{2}\,\RE(\bcE\times\bcH^*)=\frac{c_b}{2k}\,\IM(\psi^*\vec\nabla\psi),
	\label{Poynting}
	\ee
where we have made use of \eqref{TE-TM-def} and the fact that far away from the grating $\bcH=-ik\bnabla\times\bcE$.}

{Next, we recall that according to \eqref{varpi-def}, \eqref{theta-p-def}, and \eqref{sj-def},
	\begin{align}
	&\delta(k\sin\theta-k\sin\theta_{j\pm})=\frac{\delta(\theta-\theta_{j\pm})}{k|\cos(\theta)|}=
	\frac{\delta(\theta-\theta_{j\pm})}{\varpi(k\sin\theta)}~~~\for~~\pm\cos\theta>0,
	\label{id-777}
	%\\
	%&\cos\theta_{0\pm}=\pm\cos\theta_{0+}=\pm\cos\theta_0~~~\for~~\pm\cos\theta_0>0,
	%\label{id-778}\\
	%&\sin\theta_{0\pm}=\sin\theta_{0+}=\sin\theta_0.
	%\label{id-779}
	\end{align}
With the help of \eqref{asym}, \eqref{LR-waves}, \eqref{tdelta}, and \eqref{A-sum} -- \eqref{tau-jm}, and \eqref{id-777}, we can show that 	
	\begin{align}
    	\psi(x,y)&\to\frac{\sA_0}{2\pi}\Big\{e^{i\vec k_0\cdot \vec r}+ 
	\frac{i}{\sqrt{2\pi}}\sum_{j=0}^J\left(\tau_{j+} e^{ikx\cos\theta_{j+}}+
	\tau_{j-}e^{-ikx\cos\theta_{j+}}\right)e^{iky\sin\theta_{j+}}\Big\}
	~~\for~~x\to\pm\infty,\nn\\
	&\to \frac{\sA_0}{2\pi}\Big\{e^{i\vec k_0\cdot \vec r}+
	\frac{i}{\sqrt{2\pi}}\sum_{j=0}^J\big(\tau_{j+} e^{i\vec k_{j+}\cdot\vec r}+
	\tau_{j-} e^{i\vec k_{j-}\cdot\vec r}\,\big)\Big\}~~\for~~x\to\pm\infty,	
	\label{asym-neq11}
    	\end{align}
where $\vec k_{j\pm}$ are the wave vectors for the diffracted waves, i.e.,
	\be
	\vec k_{j\pm}:=k(\cos\theta_{j\pm}\,\hat\bfe_x+\sin\theta_{j\pm}\,\hat\bfe_y)
	=k(\pm\cos\theta_{j+}\,\hat\bfe_x+\sin\theta_{j+}\,\hat\bfe_y),
	\ee
and we have made use of \eqref{theta-p-def} and \eqref{sj-def}. }

{According to \eqref{asym-neq11}, the normalized diffracted beams are described by
	\be
	\psi_{j\pm}(\vec r\,):=\frac{i\tau_{j\pm}}{\sqrt{2\pi}}\, e^{i\vec k_{j\pm}\cdot\vec r}.
	\label{psi-jpm}
	\ee
For the left-incident waves, $\psi_{j+}$ and $\psi_{j-}$ respectively correspond to the transmitted and reflected diffracted beams, while the opposite is the case for the right-incident waves.}

{In view of \eqref{Poynting} and \eqref{psi-jpm}, $|\tau_{j\pm}|^2/2\pi$ give the normalized intensities of the diffracted beams. Substituting the right-hand side of \eqref{asym-neq11} in \eqref{Poynting}, we can also obtain the time-averaged Poynting vector for the total wave detected by a detector located at $\vec r$.} %

\section*{Appendix D: Expressing $\tau_{j\pm}$ in terms of 
$\sM_{j,ab}$ for $j\in\{1,2\}$}

To calculate $\tau_{1\pm}$ and $\tau_{2\pm}$, first we note that their dependence on $k$ and $\theta_0$ is independent of the range of these variables. This means that we can confine our attention to wavenumber $k<3\fK/2$ where the third- and higher-order diffracted beams are absent. For these wavenumbers, we can use \eqref{S-commutator}, \eqref{sN-ab-2},  \eqref{Rs-T-zero} , \eqref{cW-def} -- \eqref{cZ-def}, \eqref{id-324},  \eqref{id-892}, and \eqref{cN=2} to show that
	\begin{align}
	&\cN_{ab}(\theta)=-\sM_{1,ab}(k\fs_1)\Delta_1(\theta)-\sM_{2,ab}(k\fs_2)\Delta_2(\theta),
	\label{cN=C}\\
	&\begin{aligned}
	\cV(\theta)=&-\cP_{1,1}\sM_{1,12}(k\fs_2)\Delta_2(\theta),
	\end{aligned}
	\label{cXab=C}\\
	&\cW(\theta)=-\cQ_1\sM_{1,12}(k\fs_2)\Delta_2(\theta),
	\label{cW=C}\\
	&\begin{aligned}
	\cX(\theta)=&\cP_{1,1}\Delta_1(\theta)+
	(\cP_{2,2}+\cP_{1,1}\cP_{1,2})\Delta_2(\theta),
	\end{aligned}
	\label{cX=C}\\
	&\begin{aligned}
	\cY(\theta)=&\cQ_1 \Delta_1(\theta)+(\cQ_2+\cP_{1,2}\cQ_1)\Delta_2(\theta),
	\end{aligned}
	\label{cYab=C}
	\end{align}
where
	\begin{align}
	&\cP_{i,j}:=-\sM_{i,22}(k\fs_j)+\sR^l(\theta_{j+})\sM_{i,12}(k\fs_j),
	\label{cP-def}\\
	&\cQ_i:=-\sM_{i,21}(k\fs_i)+\sR^l(\theta_{i+})\sM_{i,11}(k\fs_i),
	\label{cQ-def}\\
	&\Delta_i(\theta):=\delta(\theta-\theta_{i+})+\delta(\theta-\theta_{i-}).
	\end{align}
Substituting \eqref{cN=C} -- \eqref{cYab=C} in \eqref{A-left-2} -- \eqref{B-right-2},  and comparing the resulting equations with the right-hand sides of \eqref{A-sum} and \eqref{B-sum}, we find
	\begin{align}
	&\cA_1^l=2\pi \sT(\theta_{1+})\big[
	\sM_{1,11}(k\fs_1)+\sR^l(\theta_0)\sM_{1,12}(k\fs_1)\big],
	\label{cA-L1}\\	
	&\begin{aligned}
	\cB_1^l&=\;2\pi\big[\cP_{1,1}\sR^l(\theta_0)+\cQ_1\big]\\
	&=-2\pi\big[\sM_{1,21}(k\fs_1)+\sR^l(\theta_0)\sM_{1,22}(k\fs_1)-
	\sR^l(\theta_{1+})\big\{\sM_{1,11}(k\fs_1)+
	\sR^l(\theta_0)\sM_{1,12}(k\fs_1)\big\}\big],
	\end{aligned}
	\label{cB-L1}\\
	&\cA_1^r=2\pi \sT(\theta_0)\sT(\theta_{1+}) \sM_{1,12}(k\fs_1),
	\label{cA-R1}\\
	&\cB_1^r=2\pi \cP_{1,1}\sT(\theta_0)=-2\pi\sT(\theta_0)[\sM_{1,22}(k\fs_1)-\sR^l(\theta_{1+})\sM_{1,12}(k\fs_1)],
	\label{cB-R1}\\
	&\begin{aligned}
	\cA^l_2=\;
	&2\pi\sT(\theta_{2+})\Big[\sM_{2,11}(k\fs_2)+\cQ_1\sM_{1,12}(k\fs_2)+
	\sR^l(\theta_0)\big\{\sM_{2,12}(k\fs_2)+\cP_{1,1}
	\sM_{1,12}(k\fs_2)\big\}\Big],
	\end{aligned}
	\label{AL-C}\\
	&\begin{aligned}
	\cB^l_2=2\pi\big[\cQ_2+\cP_{1,1}\cQ_1+\sR^l(\theta_0)(\cP_{2,2}+\cP_{1,1}\cP_{1,2})\big]
	\end{aligned}
	\label{BL-C}\\
	&\begin{aligned}
	\cA^r_2= &2\pi\sT(\theta_0)\sT(\theta_{2+})\Big[\sM_{2,12}(k\fs_2)+\cP_{1,1}\sM_{1,12}(k\fs_2)\Big],
	\end{aligned}
	\label{AR-C}\\
	&\begin{aligned}
	\cB^r_2= &2\pi\sT(\theta_0)(\cP_{2,2}-\cP_{1,1}\cP_{1,2}).
	\end{aligned}
	\label{BR-C}
	\end{align}
Substituting $j=1$ and \eqref{cA-L1} -- \eqref{cB-R1} in \eqref{tau-jp} and \eqref{tau-jm}, we obtain \eqref{tau-1p} and \eqref{tau-1m}. Similarly, we can express $\tau_{2\pm}$ in terms of $\sM_{1,ab}$ and $\sM_{2,ab}$ by substituting $j=2$ and \eqref{AL-C} -- \eqref{BR-C} in  \eqref{tau-jp} and \eqref{tau-jm}.

\pagebreak

\ed